\documentclass[12pt]{article}
\usepackage[totalwidth=480pt,totalheight=640pt]{geometry}
\usepackage[centertags]{amsmath}
\usepackage{array,multirow}
\usepackage{amssymb}
\usepackage{graphicx}
\usepackage{color}
\usepackage{setspace}
\pdfoutput=1

\usepackage{epsfig}

\def\m{\mu}

\def\Or[#1]{{\text{O}}\left({#1}\right)}
\def\dotl[#1,#2]{\left\langle #1, #2 \right\rangle}
\def\dotlb[#1,#2]{[ #1, #2 ]}
\def\dotp[#1,#2]{(#1) \cdot (#2)}
\def\aff[#1,#2]{\hat{#1}(#2)}
\def\n4sym{{\cal N}=4 SYM}
\def\>{\rangle}
\def\<{\langle}
\def\weight[#1,#2,#3]{\{(#1),#2,#3\}}
\def\ads[#1]{$\text{AdS}_{#1}$}

\newcommand{\be}{\begin{equation}}
\newcommand{\ee}{\end{equation}}
\newcommand{\ba}{\begin{eqnarray}}
\newcommand{\ea}{\end{eqnarray}}
\newcommand{\bea}{\begin{eqnarray}}
\newcommand{\eea}{\end{eqnarray}}

\newcommand{\CA}{{\cal A}}
\newcommand{\CS}{{\cal S}}
\newcommand{\CC}{{\cal C}}

\newcommand{\CI}{{\cal I}}
\newcommand{\CN}{{\cal N}}
\newcommand{\CO}{{\cal O}}

\newcommand{\nn}{\nonumber}

\usepackage{hyperref}
\title{Mellin Amplitudes}

\begin{document}

\begin{titlepage}

\begin{center}
\vspace{1cm}

{\Large \bf  Unitarity and the Holographic S-Matrix}

\vspace{0.8cm}

\bf{A. Liam Fitzpatrick$^1$, Jared Kaplan$^2$}

\vspace{.5cm}

{\it $^1$ Stanford Institute for Theoretical Physics, Stanford University, Stanford, CA 94305}\\
{\it $^2$ SLAC National Accelerator Laboratory, 2575 Sand Hill, Menlo Park, CA 94025, USA.} \\

\end{center}

\vspace{1cm}

\begin{abstract}

The bulk S-Matrix can be given a non-perturbative definition in terms of the flat space limit of AdS/CFT.  We show that the unitarity of the S-Matrix, ie the optical theorem, can be derived by studying the behavior of the OPE and the conformal block decomposition in the flat space limit.  When applied to perturbation theory in AdS, this gives a holographic derivation of the cutting rules for Feynman diagrams. 

To demonstrate these facts we introduce some new techniques for the analysis of conformal field theories.  Chief among these is a method for conglomerating local primary operators $\CO_1$ and $\CO_2$ to extract the contribution of an individual primary $\CO_{\Delta, \ell}$ in their OPE.  This provides a method for isolating the contribution of specific conformal blocks which we use to prove an important relation between certain conformal block coefficients and anomalous dimensions.  These techniques make essential use of the simplifications that occur when CFT correlators are expressed in terms of a Mellin amplitude.

\end{abstract}

\bigskip

\end{titlepage}

\section{Introduction}

Exact theories of quantum gravity should be formulated in terms of gauge invariant observables associated to the boundary of spacetime.  In flat spacetime, the only such observable is the S-Matrix, so a theory of quantum gravity in flat space will be a theory that computes scattering amplitudes holographically.  Since AdS/CFT \cite{Maldacena, Witten, GKP} provides a non-perturbative description of AdS theories via a dual CFT, one can obtain the bulk S-Matrix from a flat space limit of AdS.  This defines a holographic theory for flat space using a sequence of CFTs with increasing central charge.  The introduction of the Mellin amplitude \cite{Mack, MackSummary} for CFT correlation functions has led to progress \cite{Penedones:2010ue, Fitzpatrick:2011ia, Paulos:2011ie, Fitzpatrick:2011hu, Nandan:2011wc} along these lines, and in particular, we recently argued \cite{Fitzpatrick:2011hu} that bulk locality can be understood by showing how  the meromorphy of the Mellin amplitude\footnote{We reviewed aspects of CFT physics and the Mellin amplitude in our recent companion paper \cite{Fitzpatrick:2011hu}, and we discussed them in detail in \cite{Fitzpatrick:2011ia}, so we urge interested readers to consult these references for a review.} leads to an analytic S-Matrix.  The purpose of the present work is to demonstrate how the unitarity of the S-Matrix can be derived directly from the unitarity of the CFT.  Specifically, we will derive the usual optical theorem
\be
-i (T - T^\dag) = T^\dag T
\ee
and cutting rules for the 2-to-2 scattering amplitude of massless scalars at a non-perturbative level from the conformal block decomposition and the operator product expansion of the CFT.

Before we outline the derivation, let us first comment on how the standard, manifestly unitary definition of the S-Matrix can be applied in AdS/CFT.  The S-Matrix is usually defined as the overlap between in and out states
\be
\CS_{\alpha \beta} = \langle \alpha_{\mathrm{in}} | \beta_{\mathrm{out}} \rangle 
\ee
where $\alpha$ and $\beta$ are multi-particle states composed of asymptotically well-separated, exactly stable particles.  From this point of view, unitarity arises as a consequence of the completeness of the in and out bases, and all of the structure of scattering is encoded in the fact that these bases are different.  In the interaction picture we write the S-Matrix as
\be
\CS_{\alpha \beta} = \left\langle \alpha_{\mathrm{free}} \left| \CS \right| \beta_{\mathrm{free}} \right\rangle  \ \ \ \mathrm{where} \ \ \ \CS = T \left\{ e^{i \int_{-\infty}^\infty H_I(t) dt} \right\}
\label{eq:SOperator}
\ee
and $T$ is the time ordering symbol. The unitarity of the S-Matrix follows automatically from the unitarity of the $\CS$ operator.

\begin{figure}[t!]
\begin{center}
\includegraphics[width=0.95\textwidth]{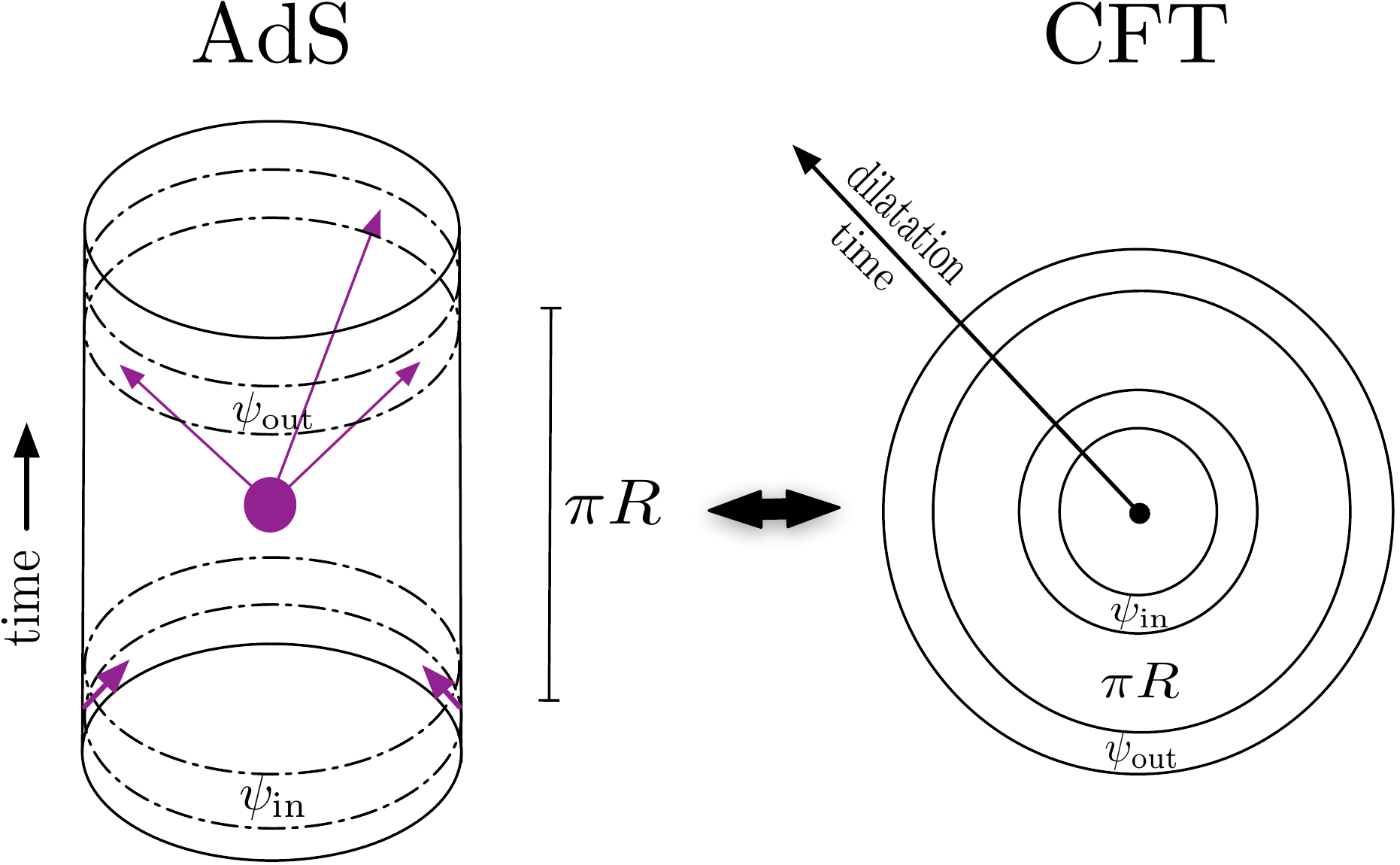}
\caption{ This figure shows how the AdS$_{d+1}$ cylinder in global coordinates corresponds to the CFT$_d$ in radial quantization.  The time translation operator in the bulk of AdS is the dilatation operator in the CFT, so energies in AdS correspond to dimensions in the CFT.  A scattering process in the bulk can be set up by acting with smeared CFT operators at an initial and final time that are separated by $\pi R$.  In the large $N$ limit, a product of $n$ single-trace CFT operators creates an $n$-particle scattering state in the bulk.
\label{fig:AdSCylinderIntro}  }
\end{center}
\end{figure}

All of these statements have simple analogs when we take the flat space limit of AdS/CFT.  The key is to realize that global AdS behaves likes a cavity or finite sized `box' \cite{Callan:1989em}, so to obtain the S-Matrix we need only setup the correct experiment and then take the size of the box to infinity.  As originally discussed in \cite{Polchinski, Susskind} and recently revisited in \cite{GGP, Fitzpatrick:2011jn, Fitzpatrick:2011hu}, one can setup initial states corresponding to incoming particles by acting with CFT operators, and then measure the outgoing particles with final state operators after exactly one scattering event has occurred.    Since we want to scatter finite energy particles in the vanishing curvature limit of AdS, we need to study bulk states with energy $E$ so that $ER \to \infty$ as the AdS length scale $R \to \infty$.  An elementary feature of the AdS/CFT correspondence is that AdS global time corresponds to radial quantization `time' in the CFT, as pictured in figure \ref{fig:AdSCylinderIntro}.  This means that time translations in the bulk of AdS are generated by the dilatation operator $D$ in the CFT, so bulk scattering amplitudes involve CFT states of dimension very large compared to $1$, but very small compared to the central charge.

In other words, to compute scattering amplitudes using AdS/CFT we setup an in-state by smearing with CFT operators at an initial ``dilatation time'', we evolve the state with the dilatation operator $D$ for a time $\pi R$, and then we measure the result at a final time.  Now it is easy to imitate the usual interaction picture.  When studying CFT operators and states with dimension small compared to the central charge $N^2$, we can separate the dilatation operator into $D = D_0 + \frac{1}{N} D_I$.  Bulk perturbation theory and bulk scattering amplitudes can be computed using equation (\ref{eq:SOperator}) with 
\be
\CS_{\alpha \beta} = \lim_{R \to \infty} \left\langle \alpha_{\mathrm{free}} \left| \CS_R \right| \beta_{\mathrm{free}} \right\rangle  \ \ \ \mathrm{where} \ \ \ \CS_R = T \left\{ \exp \left[ i \int_{-\frac{\pi R}{2} }^{ \frac{\pi R}{2} } D_I(t) dt \right] \right\}
\label{eq:SOperatorD}
\ee
where now the states $\alpha$ and $\beta$ are created by products of single trace operators, as discussed in \cite{GGP, Fitzpatrick:2011jn, Fitzpatrick:2011hu}.  From this point of view, the unitarity of the S-Matrix is a direct consequence of the unitarity of the CFT.  This description of scattering also immediately explains the S-Matrix results of \cite{Katz}, namely that to first order in perturbation theory, the bulk S-Matrix is just the matrix of anomalous dimensions $\langle \alpha | D_I | \beta \rangle$.  

While this procedure looks familiar from the point of view of the bulk, the setup of equation (\ref{eq:SOperatorD}) does not appear very natural in the CFT, nor is it convenient to use for computations.  Fortunately, in \cite{Fitzpatrick:2011hu} we proved a conjecture of Penedones \cite{Penedones:2010ue} that gives an extremely simple formula for the S-Matrix written directly in terms of the Mellin amplitude for CFT correlators.  This formula also leads to a nearly trivial relationship between the conformal block decomposition of a CFT correlator and the S-Matrix in the flat space limit.  Let us now briefly review the conformal block decomposition, which can be viewed as a consequence of unitarity in the CFT.

In any theory whatsoever, one can insert the operator $\bf 1$ as a sum over states $| \alpha \rangle \langle \alpha |$, giving
\be
\CA_4(x_i) = \sum_\alpha \langle \CO_1(x_1) \CO_2(x_2) |\alpha \rangle \langle \alpha | \CO_3(x_3) \CO_4(x_4) \rangle
\label{eq:TrivialUnitaritySum}
\ee
in the case of a 4-pt correlation function.  In theories with symmetry one can make further progress by organizing the states $| \alpha \rangle$ into irreducible representations of the symmetry group.  In flat spacetime, this means that one can use Poincar\'e invariance to break the sum into states of definite energy, invariant mass, and angular momentum and then integrate over the overall momentum of the state.  In a CFT, we can organize the states $| \alpha \rangle$ of definite dimension and angular momentum into \emph{primaries} and \emph{descendants}, where the descendant states can all be represented via actions of the translation operator $P^\mu$ on a primary state.  If we organize the sum in equation (\ref{eq:TrivialUnitaritySum}) so that all descendants are grouped together with their defining primary, we have the conformal block decomposition \cite{Belavin:1984vu, Dolan:2000ut, Dolan:2003hv, Costa:2011dw} 
\be
\CA_4(x_i) = \sum_{\Delta, \ell} P_{\Delta, \ell} B_{\Delta}^\ell(x_i)
\ee
of the CFT correlator, where the $P_{\Delta, \ell}$ are fixed numerical coefficients encoding dynamical information about the theory.  The conformal blocks $B_\Delta^\ell(x_i)$ are the universal functions that represent the contribution of a given primary and its descendants to the 4-pt correlator; these functions also depend on the dimensions $\Delta_i$  of the external operators $\CO_i$, and were recently given in Mellin space in \cite{Mack, MackSummary, Fitzpatrick:2011hu} for CFTs of arbitrary spacetime dimension.  

The conformal block decomposition can also be viewed as a consequence of the operator product expansion, and this is where its power lies.  The OPE says that we can write the product of two operators as a sum
\be
\CO_1(x_1) \CO_2(x_2) = \sum_{\Delta, \ell} c_{\Delta, \ell}^{12} b_{\Delta, \ell}^{12}(x_1,x_2,x) \CO_{\Delta, \ell}(x)
\ee
where the universal 3-pt function $ b_{\Delta, \ell}^{12}$ is fixed by conformal symmetry.  If we use the OPE twice inside a 4-pt correlator, then we can express that correlator as a sum over CFT 2-pt functions with coefficients $c_{\Delta, \ell}^{12} c_{\Delta, \ell}^{34}$.  But since local CFT operators are isomorphic to CFT states, this can also be viewed as a sum over all the states in the theory, as in the conformal block decomposition.  In other words, the OPE implies that 
\be \label{eq:IntroBlockfromOPE}
P_{\Delta, \ell} = c_{\Delta, \ell}^{12} c_{\Delta, \ell}^{34}
\ee
We have derived the well-known fact that the 3-pt correlators in a CFT in principle determine all the $n$-pt correlation functions in the theory.

\begin{figure}[t!]
\begin{center}
\includegraphics[width=0.95\textwidth]{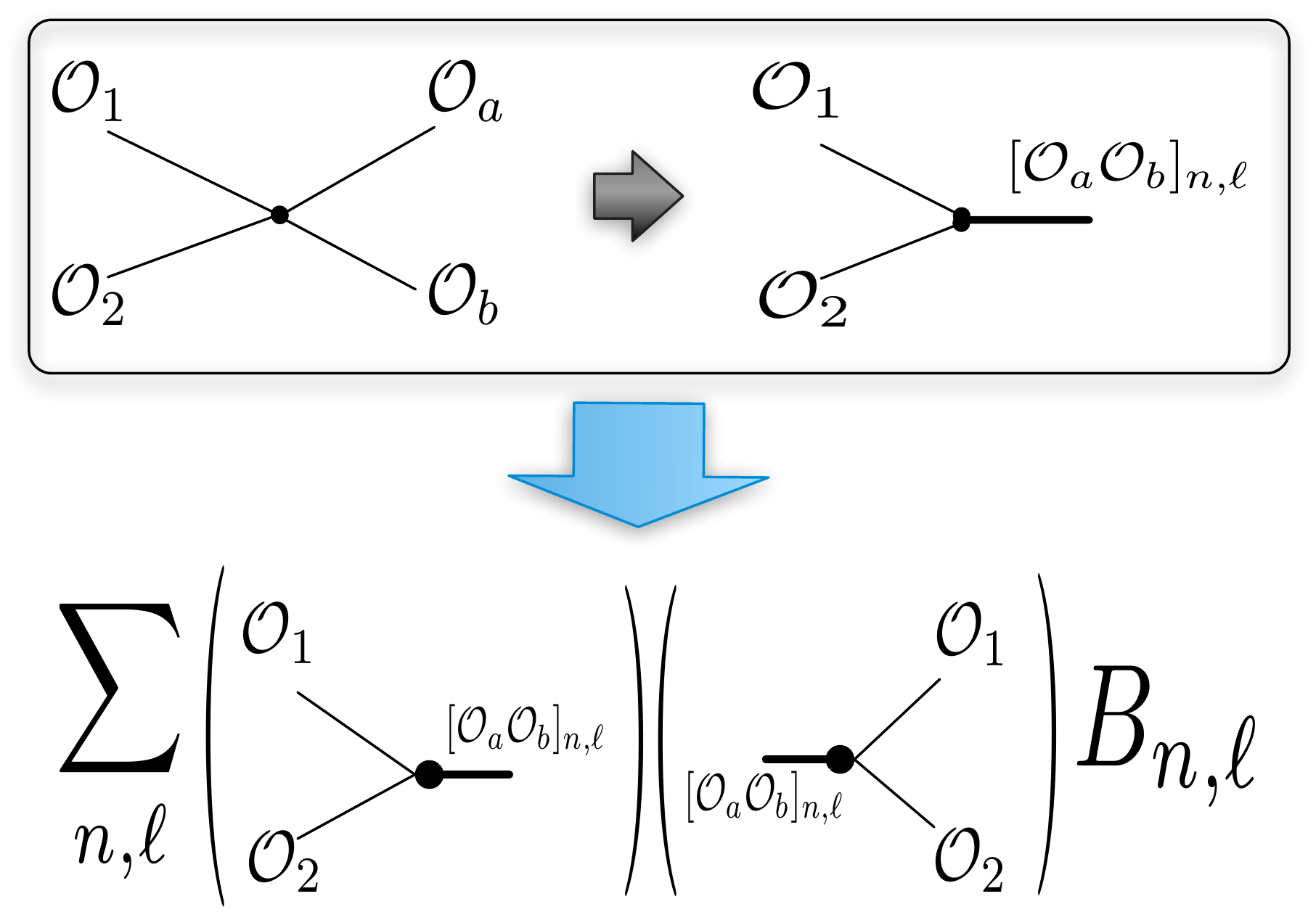}
\caption{ This figure shows how one can conglomerate $k-2$ CFT operators in an $k$-pt correlation function to obtain a 3-pt function, and then use these 3-pt functions to determine some contributions to the conformal block decomposition of a 4-point correlator.  This procedure makes it possible to use one order in perturbation theory to say something about the next; it is precisely analogous to the way that the optical theorem permits the calculation of the imaginary part of the S-Matrix using a phase space integral over the product of lower point scattering amplitudes.  
\label{fig:ConglomeratingtoBlocks}  }
\end{center}
\end{figure}

The rest of this paper will be concerned with making these ideas computationally useful and relating them to the S-Matrix.  The key to putting the OPE to work is pictured in figure \ref{fig:ConglomeratingtoBlocks};  we will refer to the process depicted in the bubble at the top of this figure as \emph{conglomerating operators} $\CO_a$ and $\CO_b$ into a double trace operator $[\CO_a \CO_b]_{n, \ell}$.  This makes it possible to use $k$-pt correlation functions to determine lower point correlators involving multi-trace operators.  In particular, we can use information about the correlators at one order in perturbation theory to compute terms in the conformal block decomposition at the next order, as pictured in the second step of figure \ref{fig:ConglomeratingtoBlocks}.  

So how do we implement this conglomeration procedure?  Naively, one might proceed by defining the double trace primary operator as a linear combination of terms of the very schematic form $\partial^x \CO_1 \partial^y \CO_2$.  Then, by imposing that the special conformal generator annihilates the sum, one finds relations for the coefficients.  By itself, this is a rather involved combinatorially exercise; some partial results were obtained in \cite{Mikhailov:2002bp, Penedones:2010ue}, and for completeness we give a recursion relation for the coefficients in the case of a general double trace primary operator in appendix \ref{app:DoubleTraceOperators}.  However, it turns out that determining these coefficients is actually the easy part, because to use these coefficients to compute correlators involving a double trace primary $[\CO_1 \CO_2]_{n, \ell}$ we also need to differentiate a CFT correlator involving $\CO_1$ and $\CO_2$ a total of $2n+ \ell$ times.  This procedure is very cumbersome, especially at large $n$ and $\ell$.

Fortunately there is a better method that exploits the simple properties of the Mellin representation for CFT correlators.  Instead of differentiating single trace operators, we can integrate them against simple `wavefunctions' that conglomerate the single-trace operators into the desired double-trace state.  To form an operator of dimension $\Delta$ and spin $\ell$ from two single trace operators inside a correlator, we write
\be
\left\langle \CO_{\Delta, \ell} (x) \dots \right\rangle = \int d^d x_1 d^d x_2 f_{\Delta, \ell} (x, x_1, x_2) \langle \CO_1(x_1) \CO_2(x_2) \dots \rangle
\label{eq:IntroConglomerate}
\ee
where `$\dots$' indicate any other local operators that may appear in the correlator.  The wavefunction $f_{\Delta, \ell}$, which we will determine in section \ref{sec:ConglomeratingOperators}, has only power law dependence on the differences between the coordinates.  Because of the simplicity of $f_{\Delta, \ell}$, when we represent CFT correlators as Mellin amplitudes, the integrals in equation (\ref{eq:IntroConglomerate}) can be done immediately using the Symanzik star formula, which one can view as the Mellin-space analog of the formula for the Fourier transform of $e^{i p \cdot x}$.  We will also see how to use these methods to extract the coefficient of an individual conformal block from the Mellin amplitude.  In section \ref{sec:ConglomeratingOperators} we will derive these techniques and use them to obtain some new results about CFTs, and then in section \ref{sec:SMatrixUnitarity} we will also make essential use of this technology in our derivation of unitarity.  It seems likely that these techniques can be usefully applied far afield from our discussion of the flat space limit of AdS/CFT.

\begin{figure}[t!]
\begin{center}
\includegraphics[width=0.95\textwidth]{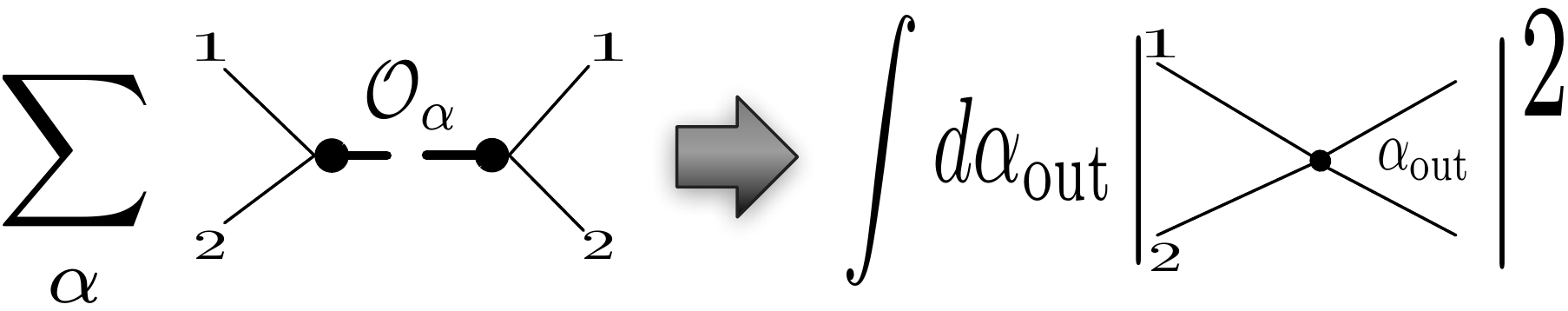}
\caption{ This figure indicates how the sum over $k$-trace operators with dimension $\Delta$ turns into a phase space integral over $k$-particle states with center of mass energy $\Delta/R$ in the flat spacetime limit of AdS/CFT.  
\label{fig:OperatorSumtoPhaseSpace}  }
\end{center}
\end{figure}

The process of conglomerating operators at one order in perturbation theory and then combining the results to give information about the next order should remind the reader of the way that the optical theorem
\be
-i (T - T^\dag) = T^\dag T 
\ee
computes the imaginary part of the S-Matrix.  In section \ref{sec:SMatrixUnitarity} we will show that in fact, the conformal block decomposition as computed along the lines of figure \ref{fig:ConglomeratingtoBlocks} reduces to the imaginary part of the S-Matrix in the flat space limit of the bulk AdS theory dual to the CFT.  

To derive the optical theorem, we need to show that the sum over $k$-trace operators in the conformal block decomposition reduces to a phase space integral over $k$-particle states, as pictured in figure \ref{fig:OperatorSumtoPhaseSpace} for $k = 2$.  We explain this essentially kinematical fact in section \ref{sec:PhaseSpace}.  One might also wonder whether all operators that can be exchanged in the conformal block decomposition are really $k$-trace operators, and what role is played by the operators dual to unstable particles.  The S-Matrix connects in and out states composed of exactly stable particles, and so scattering amplitudes between unstable particles are not well-defined and do not appear in the unitarity relation.  The qualitative difference between stable and unstable particles emerges only in the flat space limit, when the original primary operators dual to unstable particles get lost on the sea of multi-trace operators with which they mixes.  We saw an explicit example of this phenomenon in \cite{Fitzpatrick:2011hu}, where we obtained a Breit-Wigner resonance from the flat space limit of AdS/CFT.   

This also means that the small black holes that can occur as intermediate configurations in scattering processes are not literally states in the theory, since they too are unstable.  Thus there are no `small black hole operators' being exchanged in the conformal block decomposition, and we are not missing any contributions when we formulate unitarity purely in terms of stable multi-particle states.

\begin{figure}[t!]
\begin{center}
\includegraphics[width=0.95\textwidth]{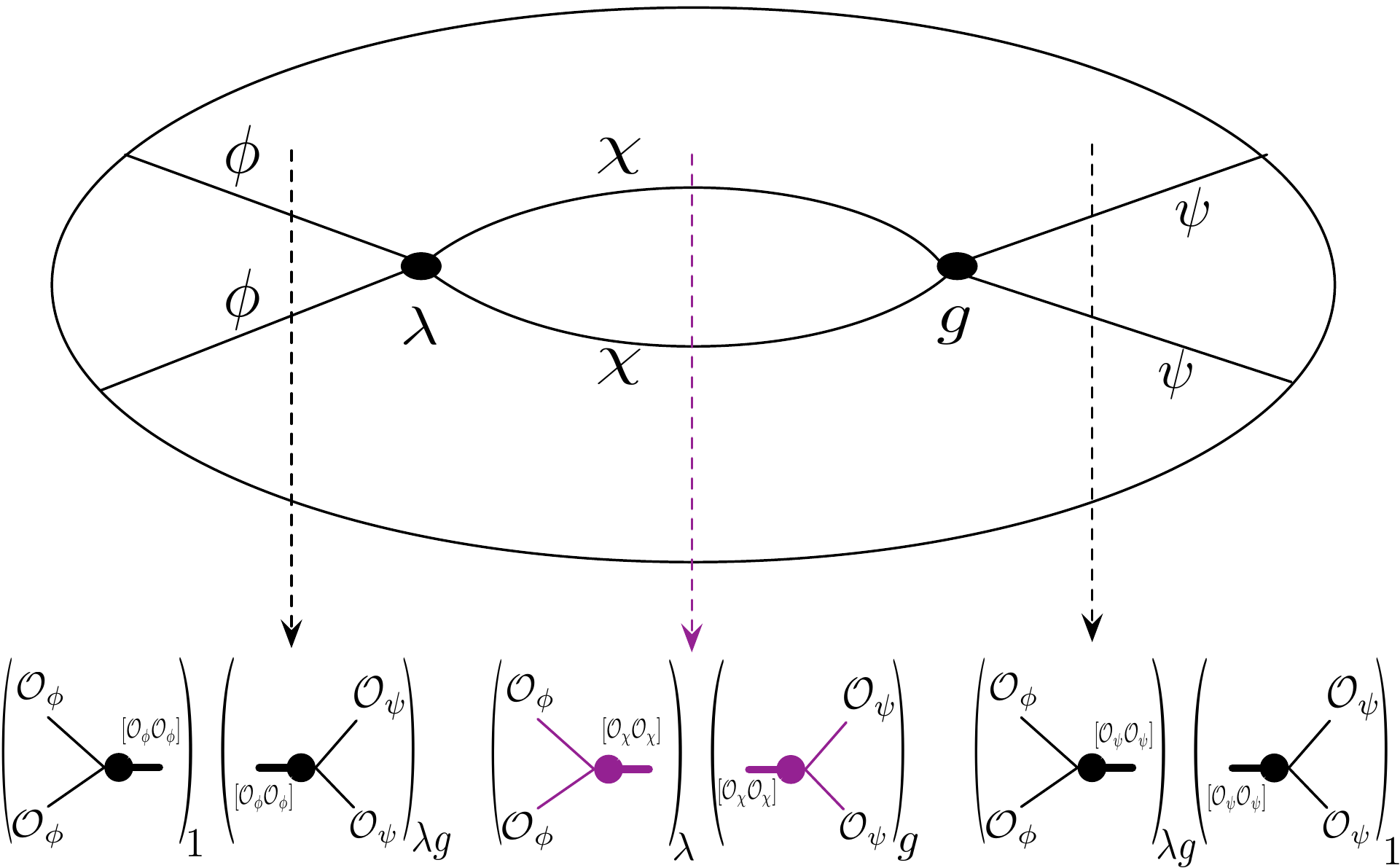}
\caption{ This figure provides a schematic depiction of how a 1-loop Witten diagram in AdS decomposes via the conformal block decomposition in the dual CFT.  For illustrative purposes, the bulk theory has both a $\frac{\lambda}{4} \phi^2 \chi^2$ and a $\frac{g}{4} \chi^2 \psi^2$ interaction.  The dashed lines indicate `cuts'; the central cut, highlighted in purple, provides the familiar imaginary contribution to the optical theorem in the flat space limit. The conformal block decomposition also includes the `edge cuts' on the left and right, which have no analog in discussions of the cutting rules.  These edge cuts are very important in order to obtain the full correlator, but in the flat space limit they only contribute to the real part of the S-Matrix, and so they drop out of the optical theorem.
\label{fig:WittenDiagramtoBlocks}  }
\end{center}
\end{figure}

The conformal block decomposition provides an expression for the exact 4-point correlator, but the left hand side of the optical theorem only involves the imaginary part of the S-Matrix.  Another way of saying this is that in general, the optical theorem does not provide sufficient information to fully determine the next order in perturbation theory, because the real part of the S-Matrix cannot be uniquely computed.  But this means that when we use the OPE as pictured in figure \ref{fig:ConglomeratingtoBlocks}, we must be missing terms that correspond to the real part of the S-Matrix!  The missing terms can be most easily understood by looking at the one-loop example in figure \ref{fig:WittenDiagramtoBlocks}.  In the conformal block decomposition of the 4-point correlator computed by this loop diagram, when double-trace operators are exchanged there are terms that correspond to `cuts at the edge of the diagram'.  These combine the 3-point functions of mean field theory (ie the CFT correlators that follow from a free theory in AdS) with interacting 3-point functions, as pictured on the left and right sides of figure \ref{fig:WittenDiagramtoBlocks}.  We will prove that in the flat space limit, these terms in the conformal block decomposition only contribute to the real part of the S-Matrix, and so they drop out of the optical theorem.  These edge cuts identically represent the part of the S-Matrix that is non-trivial to reproduce using dispersion relations.  

Interestingly, the proof that these edge cuts only contribute to the real part of the S-Matrix requires an identity first conjectured in \cite{JP}, which amounts to the statement that the edge cut terms are total derivatives.  Alternatively, the conjecture says that the OPE coefficients $c_{n,\ell} = \bar{c}_{n,\ell} + \delta c_{n,\ell}$ for these edge cuts satisfy
  \be
  \label{eq:IntroDerivativeRelation}
 \bar{c}_{n,\ell} \delta c_{n,\ell} = \frac{1}{4} \frac{\partial}{\partial n} \left(  \bar{c}_{n,\ell}^2 \gamma(n, \ell) \right),
  \ee  
  where $\gamma(n, \ell)$ is the anomalous dimension of the double trace operator $[\CO_1 \CO_2]_{n, \ell}$ of dimension $\Delta_1 + \Delta_2 + 2n + \ell + \gamma(n, \ell)$, and $\bar{c}_{n,\ell}$ and $\delta c_{n,\ell}$ are respectively the infinite $N$ value and finite $N$ corrections to $c_{n,\ell}$.  We will precisely state and prove this statement  and a relevant generalization at a non-perturbative level in section \ref{sec:FurtherApplications} by using the conglomeration techniques discussed above. 
  
The outline of the paper is as follows.  In section  \ref{sec:ConglomeratingOperators} we derive our technique for conglomerating operators and apply it to some useful examples, obtaining a few new results along the way, including the infinite $N$ conformal block coefficients in arbitrary spacetime dimensions and the generalized derivative relation indicated in equation (\ref{eq:IntroDerivativeRelation}).  In section \ref{sec:SMatrixUnitarity} we show how the unitarity of the S-Matrix follows from the conformal block decomposition, as we briefly outlined above, and we give a full one-loop example.  Finally in section \ref{sec:Discussion} we conclude with a discussion of the implications and opportunities for further work.

\section{Conglomerating Operators}
\label{sec:ConglomeratingOperators}

The goal of this section will be to understand how to `conglomerate operators' in order to combine a pair of local primary operators into a third composite operator that appears in the OPE of the first two.  We will also be able to use these techniques to extract specific terms from the conformal block expansion.

\subsection{Basics}

Before we see how the conglomeration process works, let us first review a few basic facts about CFT operators.  We have written out the full conformal algebra in equation (\ref{eq:confalgebra}) in the appendix, but for our present purposes it will suffice to consider the commutation relations of the dilatation operator $D$, the momentum generator $P^\mu$, and the special conformal generator $K^\mu$.  These take the form
\ba
\left[ D, P_\mu \right] =  P_\mu , \ \ \ \left[ D, K_\mu \right] = - K_\mu , \ \ \   \left[ P_\mu, K_\nu \right] =
- 2 (\eta_{\mu\nu} D + i M_{\mu\nu} ) 
\ea
The crucial feature to notice is that $P^\mu$ and $K^\mu$ act as raising and lowering operators with respect to the dimension, which is the eigenvalue of $D$.  Unitary CFTs have lower bounds on the allowed dimensions of operators, so this means that after some number of applications of $K^\mu$ any state of definite dimension will be annihilated.  A state that is annihilated by $K^\mu$ is called a primary state, and the operator that creates this state is referred to as a primary operator.  All states of definite dimension and angular momentum can be classified as either primaries or descendants of a primary.  Since the momentum $P^\mu = i \partial_\m$, descendants are just derivatives of primaries.

Let us begin with some very concrete examples in mean field theory (ie the dual of a free theory in AdS), where all correlators are determined by the 2-pt functions of single-trace primaries, which are the operators dual to the fields in AdS.  Given two single-trace primary scalar operators $\CO_1$ and $\CO_2$, one can form double-trace primaries $[\CO_1 \CO_2]_{n, \ell}$ which will have dimension $\Delta_1 + \Delta_2 + 2n + \ell$  and spin $\ell$.  The most trivial example is $\CO_1 \CO_2$, which is primary and has $n = \ell = 0$.  But we can also form the operator
\be
\Delta_2 ( \partial_\mu \CO_1 ) \CO_2 - \Delta_1 \CO_1 (\partial_\mu \CO_2)
\ee
One can check using the conformal algebra that this operator is primary.  A slightly more complicated example is the operator given by the linear combination
\be
\frac{\Delta_1}{2 \Delta_1 + 2 - d} ( \partial^2 \CO_1 ) \CO_2 -  \partial_\mu \CO_1 \partial^\mu \CO_2 + \frac{\Delta_2}{2 \Delta_2 + 2 - d} \CO_1 (\partial^2 \CO_2)
\ee
which is also primary, and has $n = 1, \ell = 0$.  In appendix \ref{app:DoubleTraceOperators}  we present a recursion relation that completely determines the appropriate coefficients for any double-trace primary with arbitrary $n$ and $\ell$.  At large $n$, the double trace primary operators approach a one-to-one correspondence with the space of 2-particle states in $d+1$ dimensions, a fact that will be important later on.

A very natural question follows: given an n-pt CFT correlator involving $\CO_1(x_1)$ and $\CO_2(x_2)$, how do we extract an $(n-1)$-pt correlator with the double trace primary $[\CO_1 \CO_2]_{n, \ell}(x_0)$?  We could proceed by using derivatives as above, but this quickly becomes extremely cumbersome.  Furthermore, when we move beyond the mean field theory limit, all operators will pick up anomalous dimensions, and when these become large it is difficult to define precisely which operator we intend when we write $[\CO_1 \CO_2]_{n, \ell}(x_0)$.   So instead of using derivatives, let us try to use an integral over a wavefunction $f_{\Delta, \ell}$, and define 
\be \label{eq:WarmUpWavefunctionDesired}
[\CO_1 \CO_2]_{n, \ell}(x_0) = \int d^d y_1 d^d y_2 f_{\Delta_1 + \Delta_2 + 2n+ \ell, \ell}(x_0, y_1, y_2) \CO_1(y_1) \CO_2(y_2)
\ee
This applies to the mean field theory case, but for general CFTs we can use dimensions $\Delta$ other than $\Delta_1 + \Delta_2 + 2n + \ell$.  Now we need to determine the wavefunction $f_{\Delta, \ell}$. Fortunately, this can be easily accomplished with the introduction of so-called shadow operators.  For any primary scalar operator $\CO$ of dimension $\Delta$, we wish to find a shadow operator $\tilde \CO$ of dimension $d - \Delta$ so that
\be
\langle \CO(x) \tilde \CO(y) \rangle = \delta^d(x-y)
\ee
If we simply define
\be
\tilde \CO(x) \equiv \int d^dy \frac{\CO(y)}{(x-y)^{d - 2 \Delta}}
\ee
then one can check that the desired identity is satisfied.  Now we can use these shadow operatos to compute the wavefunction $f_{\Delta, \ell}$.  As a warm-up, assume that $\ell = 0$.  If we compute the correlator of both sides of equation (\ref{eq:WarmUpWavefunctionDesired}) with $\tilde \CO_1(x_1) \tilde \CO_2(x_2)$ then we find that 
\be
f_{\Delta, \ell=0}(x_0, x_1, x_2) \propto \frac{(x_{12})^{\Delta_1 + \Delta_2 - 2d + \Delta } }{(x_{01})^{\Delta - \Delta_1 +\Delta_2} (x_{02})^{\Delta-\Delta_2+ \Delta_1} } 
\ee
where $x_{ij} = x_i - x_j$.  We did not even need to do any integrals to compute this wavefunction, because the 3-pt correlator of scalar primaries is determined uniquely up to an overall constant.

Before we make use of this result, let us first generalize it to the case where $\ell > 0$.  For this purpose, it will be simpler and more elegant to use the embedding formalism \cite{ Dirac:1936fq, Weinberg:2010fx}, where we can use the machinery that was developed and nicely explained in \cite{Costa:2011mg}.  The basic idea of this formalism is extremely simple -- since the conformal group in $d$ dimensions is $SO(d, 2)$, it is most natural to use coordinates that transform in the fundamental representation of this group.  Thus we will represent each coordinate $x_i$ with a $d+2$ dimensional vector $P_i$, constrained so that $P_i^2 = 0$ and identified projectively so that $P_i \sim \lambda P_i$ for real $\lambda > 0$.   These coordinates correspond to the null cone that is the asymptotic limit and boundary of AdS when it is regarded as a hyperbola in a $d+2$ dimensional embedding space.  If we use light cone coordinates for the $P_i$ and choose the specific normalization $P_i^+ = 1$, we find
\be
(P_i^+, P_i^-, P_i^\mu) = (1, x_i^2, x_i^\mu)
\ee
This means that the inner product of the $P_i$ is
\be
2 P_i \cdot P_j = P_i^+ P_j^- + P_i^- P_j^+ - P_i^\mu P_{j \mu} = (x_i - x_j)^2
\ee
Conformal transformations of the $x_i$ simply act as fundamental $SO(d,2)$ transformations on the $P_i$.  We will normalize the 2-pt functions of single-trace scalar primary operators so that
\be
\langle \CO(P_1) \CO(P_2) \rangle = \frac{\CC_\Delta}{P_{12}^{\Delta}}
\ee
where $P_{12} = 2 P_1 \cdot P_2$ and the normalization
\be
\CC_\Delta \equiv \frac{\Gamma(\Delta)}{2 \pi^h \Gamma(\Delta - h + 1)}
\ee
with $2h = d$, the spacetime dimension of the CFT.  As shown in \cite{Costa:2011mg}, the correlators of operators with spin can be described in the embedding formalism as polynomials in auxiliary $d+2$ dimensional vectors $Z_i$ which soak up the tensor indices of the spinning operators.  We will only make use of the simplest examples from \cite{Costa:2011mg}, such as the 3-pt function of two scalar primaries and a spin $\ell$ primary
\ba \label{eq:ThreePointFunction}
Z_{A_1} ... Z_{A_\ell} \left\langle \CO_1 (P_1) \CO_2 (P_2) \CO_{\Delta, \ell}^{A_1... A_\ell} (P_3) \right\rangle
&=& \left(c^{12}_{\Delta, \ell} \right) T_{ \Delta_1, \Delta_2}^{\Delta, \ell}(Z, P_3; P_1, P_2 )
\\
\mathrm{where} \ \ \ T_{ \Delta_1, \Delta_2}^{\Delta, \ell}( Z, P_3; P_1, P_2  ) & \equiv & \frac{\left((Z \cdot P_1) P_{23} - (Z \cdot P_2) P_{13} \right)^\ell}{P_{12}^{\frac{\Delta_1 + \Delta_2 - \Delta + \ell}{2}} P_{23}^{\frac{\Delta_2 + \Delta - \Delta_1 + \ell}{2}} P_{31}^{\frac{\Delta + \Delta_1 - \Delta_2 + \ell}{2}} }
\ea
Note that the universal function $T_{ \Delta_1, \Delta_2}^{\Delta, \ell}$ is fixed by symmetry, while the 3-pt function coefficient $c_{\Delta, \ell}^{12}$ provides dynamical information about the theory.  The auxiliary coordinates $Z_i$ are taken to have the property that $Z_i \cdot P_i = 0$, and correlators must have a `gauge invariance' under $Z_i \to Z_i + \alpha P_i$ for any $\alpha$.  One can see immediately that the scalar-scalar-spin-$\ell$ correlator satisfies this gauge condition.

We can use these results to determine the general wavefunction $f_{\Delta, \ell}$.  The operator $\CO_{\Delta, \ell}$ is defined by
\be \label{eq:DefinitionConglomeration}
Z_{A_1}... Z_{A_\ell}  \CO_{\Delta, \ell}^{A_1... A_\ell}(P_0) = \int d^d P_1 d^d P_2 \left[ Z_{A_1}... Z_{A_\ell} f_{\Delta, \ell}^{A_1... A_\ell}(P_0, P_1, P_2) \right] \CO_1(P_1) \CO_2(P_2)
\ee
If we again take the correlator of both sides with the product of shadow fields $\tilde \CO_1(P_1) \tilde \CO_2(P_2)$ then we find the result
\be \label{eq:ConglomeratingWavefunction}
Z_{A_1} \dots Z_{A_\ell} f^{A_1, \dots, A_\ell}_{\Delta, \ell}( P_0, P_1, P_2) = \frac{1}{N_{\Delta, \ell}^f} T_{d - \Delta_1, d - \Delta_2}^{\Delta, \ell}(Z, P_0; P_1, P_2) 
\ee
defined in terms of the universal function from equation (\ref{eq:ThreePointFunction}), where $N^f_{\Delta, \ell}$ is a normalization factor that we will determine later.  Note that this wavefunction also depends on the spacetime dimension $d$ and the dimensions $\Delta_1$ and $\Delta_2$, although we have suppressed this dependence in the notation.  One could continue on and use the results of \cite{Costa:2011mg} to obtain wavefunctions involving several operators with spin, but for our purposes equation (\ref{eq:ConglomeratingWavefunction}) will be sufficient.  Now let us see how to use this result to compute interesting 3-pt functions and to extract the coefficients in the conformal block decomposition.

\subsection{Using Conglomeration}
\label{sec:UsingConglomeration}

The intuition we used above to introduce conglomeration was perturbative.  In general, our conglomeration procedure can be understood in terms of the operator product expansion.  Isolating a single term in the OPE, we can write
\be
\CO_1(P_1) \CO_2(P_2) = c^{12}_{\Delta, \ell; \rm OPE} b_{\Delta, \ell}^{12}(P_1,P_2)   \CO_{\Delta, \ell}(P_1) + \ldots 
\ee
where the `$\ldots$' contain all the other operators in the OPE, including the descendants of $\CO_{\Delta, \ell}$.   If we compute the correlator of both sides with $\CO_{\Delta, \ell}(P_0)$ all of the terms in the ellipsis vanish, and we relate the three-point correlator to the OPE coefficient multiplied by the normalization of the OPE and the operator $\CO_{\Delta, \ell}$.
For example, in the notationally simple case $\ell = 0$, we define $ b_{\Delta, 0}^{12} = P_{12}^{-\Delta_b + \frac{\Delta}{2}}$ and so we find that
\be
\langle \CO_{\Delta, 0}(P_0) \CO_1(P_1) \CO_2(P_2) \rangle = c^{12}_{\Delta, \ell; \rm OPE} P_{12}^{-\Delta_b + \frac{\Delta}{2}} \langle \CO_{\Delta, 0}(P_0) \CO_{\Delta, 0}(P_1) \rangle .
\ee
When $\CO_{\Delta, 0}$ is normalized to give a two-point function $P_{01}^{-\Delta}$, then we have
\be
c_{\Delta, \ell}^{12} = c^{12}_{\Delta, \ell; \rm OPE}
\ee
using the definition of the 3-point correlator in equation (\ref{eq:ThreePointFunction}).  Conglomeration makes it possible to extract both 3-point correlators and OPE coefficients.

Before proceeding to calculate we need to set the normalization.  We will use a convention such that the general 2-point correlator of $\CO_{\Delta, \ell}$ with itself is \cite{Costa:2011mg} 
\be
\left\langle \CO_{\Delta, \ell}(Z_1, P_1) \CO_{\Delta, \ell}(Z_2, P_2) \right\rangle 
=  \frac{((Z_1 \cdot Z_2) (P_1 \cdot P_2) - (Z_2 \cdot P_1) (Z_1 \cdot P_2) )^\ell}{P_{12}^{\Delta_1 + \Delta_2 + 2n + 2 \ell}},
\ee
where we have effectively defined the operator  $\CO_{\Delta, \ell}$ by conglomeration in equation (\ref{eq:DefinitionConglomeration}).  Now we can determine the normalization of the wavefunctions $N^f_{\Delta, \ell}$ in equation (\ref{eq:ConglomeratingWavefunction}) by demanding
\be \label{eq:NormalizingCorrelator}
\left\langle \CO_{\Delta, \ell} \CO_{\Delta, \ell} \right\rangle  = \int d^d P_i  f_{\Delta, \ell}(P_a; P_1, P_2) f_{\Delta, \ell}(P_b; P_3, P_4)  \left\langle \CO_1(P_1) \CO_2(P_2) \CO_1(P_3) \CO_2(P_4) \right\rangle   ,
\ee
where we have suppressed the dependence of $f_{\Delta, \ell}$ on the auxiliary variables $Z_i$ for notational simplicity.  One could also set the normalizations in terms of the conformal block decomposition.  This follows because
\be
\int d^d P_1 d^d P_2 f_{\Delta, \ell}(P_0; P_1, P_2) B_{\Delta', \ell'}(P_i) \propto  \delta( \Delta - \Delta') \delta_{\ell, \ell'} .
\ee
If the result is non-vanishing it can be used to fix the normalization of the wavefunctions; this also means that we can use conglomeration to uniquely identify terms in the conformal block decomposition.

Finally, let us consider what happens if there is more than one operator with the dimension, angular momentum, and global charges of $\CO_{\Delta, \ell}$.  If there are many such operators, they can certainly mix with each other, so they can only be differentiated based on their correlation functions.  By applying  conglomeration to different correlators, such as
\be
\left\langle \CO_1 \CO_2 \CO_1 \CO_2 \right\rangle, \ 
\left\langle \CO_1 \CO_2 \CO_3 \CO_4 \right\rangle, \
\left\langle \CO_3 \CO_4 \CO_3 \CO_4 \right\rangle   
\ee
we can extract all the information we need to separate the operator $\CO_{\Delta, \ell}$ that couples to $\CO_1$ and $\CO_2$ from the operator $\CO_{\Delta, \ell}'$ which has a 3-pt function with $\CO_3$ and $\CO_4$.  This may be relevant for more complicated CFTs.

Now that our wavefunctions are normalized, we can compute 3-point correlators via
\be
\int d^d P_1 d^d P_2 f_{\Delta, \ell}(P_0; P_1, P_2)  \left\langle \CO_1(P_1) \CO_2(P_2) \CO_3(P_3) \CO_4(P_4) \right\rangle
= \left\langle \CO_{\Delta, \ell}(P_0) \CO_3(P_3) \CO_4(P_4) \right\rangle .
\ee
  If there is no operator or operators $\CO_{\Delta, \ell}$ in the OPE of $\CO_1$ and $\CO_2$ then the result will be zero.  This procedure is most tractable when the correlators are expressed in terms of the Mellin amplitude $M(\delta_{ij})$ \cite{Mack, MackSummary, Penedones:2010ue, Fitzpatrick:2011ia, Fitzpatrick:2011hu}, so that
\be
\left\langle \CO_1(P_1) \CO_2(P_2) \CO_3(P_3) \CO_4(P_4) \right\rangle = \int_{- i \infty}^{i \infty} [ d \delta ] M(\delta_{ij}) \prod_{i < j}^4 \Gamma(\delta_{ij}) P_{ij}^{- \delta_{ij}} .
\ee
The reason is that the only dependence on the $P_{ij}$ is in the form of a power-law, and the projective integrals over the $P_i$ that we find when we conglomerate can be easily accomplished using the Symanzik star formula, which states that
\be \label{eq:SymanzikStar}
\int d^d P \prod_{i=1}^n \Gamma(l_i) (-2P_i\cdot P)^{-l_i}
=\pi^h \int  [d \delta]\, 
\prod_{i<j}^n \Gamma( \delta_{ij}) P_{ij}^{-\delta_{ij}} 
\ee
where the $\delta_{ij}$ integration variables are constrained by $\sum_{i \neq j}^n \delta_{ij} = l_i$.  Note that when $n = 3$ this means that the Mellin space integration variables $\delta_{ij}$ on the right hand side are completely fixed, so there are no integrals to do.   One can think of this very useful formula as the analog of the Fourier transform of $e^{i p \cdot x}$ in momentum space.

We will first make use of this technology to determine the conformal block decomposition of mean field theory in any number of dimensions.  This simple result has been obtained for $d = 2$ and $d=4$ in \cite{JP}, but we are not aware of it appearing anywhere in the literature for the case of general $d$.  The relevant correlator is simply
\be
\left\langle \CO_1(P_1) \CO_2(P_2) \CO_1(P_3) \CO_2(P_4) \right\rangle   = \frac{\CC_{\Delta_1} \CC_{\Delta_2} }{P_{13}^{\Delta_1} P_{24}^{\Delta_2} }
\ee
To extract the conformal block decomposition for $\ell = 0$ and $\Delta = \Delta_1 + \Delta_2 + 2n$, we simply need to integrate
\be
\int d^d P_3 d^d P_4 \frac{1}{P_{34}^{\frac{2d-\Delta_1-\Delta_2 - \Delta}{2}} P_{04}^{\frac{\Delta_1 + \Delta - \Delta_2}{2}} P_{03}^{\frac{\Delta + \Delta_2 - \Delta_1}{2}} } \times \frac{\CC_{\Delta_1} \CC_{\Delta_2}}{P_{13}^{\Delta_1} P_{24}^{\Delta_2} } .
\ee
We can apply the Symanzik star formula of equation (\ref{eq:SymanzikStar}) to the integrals over $P_1$ and $P_2$.  In both cases the constraints completely determine the integrals over the $\delta_{ij}$, and we find
\ba
\pi^{2h} \frac{\Gamma (-n) \Gamma \left(h-\Delta_1\right) \Gamma
   \left(h-\Delta_2\right) \Gamma
   \left(-h+n+\Delta_1+\Delta_2\right)}{\Gamma
   (\Delta_1) \Gamma (\Delta_2) \Gamma
   \left(h+n\right) \Gamma (2h-n-\Delta_1-\Delta_2)} 
\left( \CC_{\Delta_1} \CC_{\Delta_2}    \frac{P_{12}^n}{P_{01}^{\Delta_1 + n} P_{02}^{\Delta_2 + n } } \right),
\label{eq:scalarc2}
\ea
where as usual $2h = d$.  The coefficient outside the parentheses is $\bar c^{12}_{\Delta, 0} N^f_{\Delta, 0}$.  

In the limit that $n$ approaches a non-negative integer, the above expression is singular. Ultimately, we are interested in extracting double-trace operators whose dimensions are exactly given by integer $n$, and thus one might be concerned about whether we are really able to regulate this singularity.  For the reader
%named Joao Penedones
 who is interested in such subtleties, we will show in Appendix \ref{sec:congscalar} the details of how we choose our regulator.  For the more casual reader, however, the idea of the following derivation is relatively simple: once we look at the physically normalized operator, the singularity in the three-point function cancels against a singularity in the normalization factor, so that the physical three-point function is finite. In practice in the following, this cancellation of singularities will take the form $\Gamma(-n)/\Gamma(0)$, which we will take to be $\frac{(-1)^n}{n!}$.\footnote{Briefly, one can precisely regulate the $\Gamma(0)$ and $\Gamma(-n)$ singularities by taking the dimensions $\Delta$,$\Delta'$ of the conglomerated operators $[\CO_1 \CO_2]_{\Delta,0}$ and $[\CO_3 \CO_4]_{\Delta',0}$ to differ until physical quantities are calculated.   } To fix the normalization factor, we can conglomerate again to compute the 2-pt function of $[\CO_1 \CO_2]_{n, 0}$.

In fact, it is worthwhile to pause and note that beginning with any 3-pt function we can conglomerate $\CO_1$ and $\CO_2$ to obtain a 2-pt function.  This relates the coefficient $c_{\Delta, 0}^{12}$ that sets the size of the 3-point correlation to a coefficient $c_2^{\Delta, \ell}$ in a 2-point correlator.  To compute the relation, we multiply the 3-pt function by $f_{\Delta, \ell}$ and integrating over $P_1$ and $P_2$ to obtain the 2-pt function
\be 
c_2^{\Delta,0} = \frac{c^{12}_{\Delta,0}}{N^f_{\Delta, \ell}} \frac{\pi^{2h} \Gamma(0)\Gamma \left(h- \frac{\Delta+\Delta_1-\Delta_2}{2} \right) \Gamma(\Delta -h) \Gamma \left(h- \frac{\Delta+\Delta_2-\Delta_1}{2} \right) }{\Gamma(h) \Gamma \left(\frac{\Delta+\Delta_1-\Delta_2}{2} \right) \Gamma \left(\frac{\Delta+\Delta_2-\Delta_1}{2} \right) \Gamma(d-\Delta)}.
\label{eq:c2overc3}
\ee
We see that there is again a singularity of the form ``$\Gamma(0)$".  If the correlator that we are computing is of the form in equation (\ref{eq:NormalizingCorrelator}), then we must have $c_2^{\Delta, \ell}  = 1$, in which case we find the very useful fact
\be \label{eq:SpinZeroNorm}
N^f_{\Delta, 0} = c^{12}_{\Delta,0} \frac{\pi^{2h} \Gamma(0)\Gamma \left(h- \frac{\Delta+\Delta_1-\Delta_2}{2} \right) \Gamma(\Delta -h) \Gamma \left(h- \frac{\Delta+\Delta_2-\Delta_1}{2} \right) }{\Gamma(h) \Gamma \left(\frac{\Delta+\Delta_1-\Delta_2}{2} \right) \Gamma \left(\frac{\Delta+\Delta_2-\Delta_1}{2} \right) \Gamma(d-\Delta)} .
\ee
Note that this is a non-perturbative result, and is not restricted to mean field theory.  

Now we can incorporate the normalization $N^f_{\Delta, 0}$ and compute the desired conformal block coefficient for mean field theory.  As we discussed near equation (\ref{eq:IntroBlockfromOPE}), the coefficient is simply
 \ba
\left(\bar c^{12}_{n,0} \right)^2  &=&  \CC_{\Delta_1} \CC_{\Delta_2} 
\frac{(\Delta_1)_n (\Delta_2)_n \left(1 + \Delta_1-h \right)_n
  \left(1+\Delta_2-h \right)_n}{n! 
   \left(h \right)_n  (\Delta_1+\Delta_2 - 2h +1+n)_n  \left(\Delta_1+\Delta_2 - h+ n\right)_n} ,
   \label{eq:scalarOPEinfN}
 \ea
 where we recall that the Pochhammer symbol $(a)_b = \Gamma(a+b)/\Gamma(a)$, and the spacetime dimension in the CFT is $2h$.  In appendix \ref{sec:congspin}, we generalize this method to arbitrary spin conformal blocks in the scalar four-point function.  The integrals can also be performed in this case, with just a bit more book-keeping to track the various terms that appear when we expand the degree $\ell$ polynomial in $f_{\Delta, \ell}$.  This gives the resulting compact form for the conformal block coefficients in mean field theory (ie a free scalar theory in AdS, or a CFT at infinite $N$):
\ba
\left(\bar c^{12}_{n,\ell} \right)^2    &=& \frac{ \CC_{\Delta_1} \CC_{\Delta_2}  (-1)^\ell (\Delta_1-h+1)_n (\Delta_2-h+1)_n
   (\Delta_1)_{\ell+n} (\Delta_2)_{\ell+n}}{\ell! n! (\ell+h)_n (\Delta_1+\Delta_2+n-2h+1)_n (\Delta_1+\Delta_2+2 n+\ell-1)_l (\Delta_1+\Delta_2+n+\ell-h)_n} .\nonumber \\
   \ea
This result matches that of \cite{JP} in the cases they considered, namely that of 2-dimensional and 4-dimensional CFTs with $\Delta_1 = \Delta_2$ normalized without the factor of $\CC_{\Delta_1} \CC_{\Delta_2}$.

\subsection{Further Applications}
\label{sec:FurtherApplications}

While extracting the OPE coefficients of double-trace operators in the infinite $N$ theory is a useful example of conglomeration, its full power comes from the fact that it is an essentially non-perturbative technique, and one can use it to extract the coefficient of an arbitrary operator in the OPE.  Thus, if we know all the four-point functions of a set of operators, $\CO_1, \CO_2, \CO_3, \CO_4$, then we can conglomerate $\CO_1 \CO_2$ to make an operator $[\CO_1 \CO_2]_{\Delta, \ell}$ of arbitrary dimension $\Delta$ and spin $\ell$, with $\Delta$ a free parameter.  We do not need to know $\Delta$ a priori.  Rather, the result of conglomerating will give vanishing OPE coefficients except at the values of $\Delta$ for which there actually is a corresponding operator in the $\CO_1 \CO_2$ OPE.  We will now turn to examples where we look at the connected four-point functions of the theory and use conglomerating methods to extract information about specific conformal blocks.  Mellin space is an essential tool in this study, since by construction it organizes the connected correlators into contributions of definite powers of the $P_{ij}$'s.  Correlators can then be integrated against the wavefunctions simply by repeated application of Symanzik's star formula.  

In this subsection, we will first discuss general results on the application of the wavefunctions to connected four-point functions, in particular how to extract both OPE coefficients and anomalous dimensions.  We will then turn to the application of these results to specific AdS models.    A direct consequence of our methods will be the proof of an important derivative relation, discovered empirically in \cite{JP}, between OPE coefficients and anomalous dimensions of double-trace conformal blocks:
\ba
\bar c^{12}_{n, \ell} \delta c^{12}_{n,\ell} &=& \frac{1}{4} \frac{\partial}{\partial n} \left( \left(\bar c^{12}_{n,\ell} \right)^2 \gamma(n) \right) ,
\label{eq:derivreln}
\ea
where $\bar c^{12}_{n,\ell}$ are the infinite $N$ OPE coefficients of double-trace conformal blocks and $\delta c^{12}_{n,\ell} $ are the differences between the exact OPE coefficients and the infinite $N$ OPE coefficients.  In general, this formula is true only to leading order in perturbation theory, but as we will explain in detail, for a certain class of contributions it actually holds exactly.  

\subsubsection{OPE Coefficients from Connected Diagrams}

Consider the Mellin amplitude for a four-point function  $\langle \CO_1 \CO_2 \CO_3 \CO_4 \rangle$, and let $\Delta_a = \Delta_1 = \Delta_2$ while $\Delta_b = \Delta_3 = \Delta_4$.  The four-point function has only two independent Mellin variables, which we can choose to be $\delta \equiv 2(\Delta_a -\delta_{12}), \gamma \equiv 2( \delta_{14} +\delta_{12} - \Delta_a) $
\ba
\CA_4(P_i) = \< \CO_1(P_1) \CO_2(P_2) \CO_3(P_3) \CO_4(P_4) \> &\rightarrow & M_{1234}(\delta, \gamma). 
\ea
One can think of $\delta$ in analogy with the mandelstam invariant $s$, and so when we look in the $s$-channel, the angular momentum information will be carried by the $\gamma$ variable.  Conglomerating $\CO_1 \CO_2$ to produce $[\CO_1 \CO_2]_{\Delta, \ell}$ involves integrating the correlation function against our wavefunction $f_{\Delta, \ell}$.  Mellin space is ideally suited for this integration, since its form is already a decomposition of the correlator into powers of $P_{ij}$'s, for which the wavefunction integrations just involve a repeated use of Symanzik's star formula.  Thus, one obtains a general formula for the three-point function of $ [\CO_1 \CO_2]_{\Delta, \ell}$ with $\CO_3, \CO_4$, of the form in equation (\ref{eq:ThreePointFunction}) with coefficient
\ba
c^{34}_{\Delta, \ell} &=& \int \frac{d\delta d \gamma}{(2\pi i)^2} M_{1234}(\delta, \gamma) H_{\Delta, \ell}(\delta, \gamma),
\ea
$[\CO_1 \CO_2]_{\Delta , \ell}$ defined this way still has to be normalized, for which one must calculate its two-point function as we discussed in the $\ell = 0$ case near equation (\ref{eq:SpinZeroNorm}).  In general this is singular, and must be regulated by instead calculating the two-point function $\< [\CO_1 \CO_2]_{\Delta, \ell} [\CO_1 \CO_2]_{\Delta', \ell}\> $ with $\Delta \ne \Delta'$.  One then takes $\Delta \rightarrow \Delta'$ at the end of the calculation, and then only in physically normalized three-point function coefficients.  In order for the result to be non-zero, we must have an $\infty/\infty$ behavior as $\Delta$ approaches its physical value.  We will see shortly that this occurs only at values of $\Delta$ for which there is a pole in the Mellin integrand, as we should expect on the general grounds discussed in \cite{Fitzpatrick:2011ia, Fitzpatrick:2011hu}.  

To avoid unwieldy formulae, we will focus on the special case where the connected four-point function in question contains only conformal blocks of spin-0, for instance corresponding to s-channel scalar exchange in AdS, and leave the general case to appendix \ref{app:General4ptConglomeration}.  Then, the Mellin amplitude does not depend on $\gamma$, and the four-point amplitude takes the form
\ba
\CA_4(P_i)  &=& \int \frac{d\delta d\gamma}{(2\pi i)^2} \frac{(-1)}{4} \frac{M(\delta) \Gamma(\Delta_a -\frac{\delta}{2}) \Gamma(\Delta_b - \frac{\delta}{2}) \Gamma^2(-\frac{\gamma}{2}) \Gamma^2(\frac{\delta+\gamma}{2})}{P_{12}^{\Delta_a-\frac{\delta}{2}}
P_{34}^{\Delta_b-\frac{\delta}{2}} (P_{13} P_{24})^{-\frac{\gamma}{2}} (P_{14}P_{23})^{\frac{\delta+\gamma}{2}}} .
\ea
We can obtain from this the $\<[\CO_1 \CO_2]_{\Delta, 0} \CO_3 \CO_4 \>$ three-point function by conglomerating $\CO_1$ and $\CO_2$ together:
\ba
\CA_3(P_i)  &=& \frac{1}{N^{f12}_{\Delta, 0}} \int \frac{d P_1 dP_2}{P_{12}^{d-\Delta_a - \frac{\Delta}{2} } P_{01}^{\frac{\Delta}{2} } P_{02}^{\frac{\Delta}{2}}} \CA_4(P_i).
\ea
This may be evaluated by applying Symanzik's integral twice, which introduces two new Mellin variables (two from the $P_1$ integration, and none from the
$P_2$ integration).  However, three of these integrations are purely kinematic, in that $M$ does not depend on them, and so can be done
independently of $M(\delta)$.  Fortunately, performing first the $d \gamma$ integration, all three of them take the form of Barnes' Lemmas, and can be computed in closed form.  We arrive at 
\ba
\CA_3(P_i)  &=& \frac{(-1)\pi^{2h}}{4 N^{f12}_{\Delta, 0}} \frac{\Gamma^2(h-\frac{\Delta}{2})\Gamma^2(\frac{\Delta}{2})}{\Gamma(h)\Gamma(2h-\Delta)\Gamma(\Delta)}
 \frac{1}{P_{03}^{\frac{\Delta}{2}} P_{04}^{\frac{\Delta}{2}} P_{34}^{-\frac{\Delta}{2}+\Delta_b}} \nn\\
    && \times
     \int \frac{d \delta}{2\pi i}  \Gamma\left( \frac{\delta}{2}-\frac{\Delta}{2}\right) \Gamma\left(\Delta_a-\frac{\delta}{2}\right)
 \Gamma\left(\frac{\delta}{2} -h+\frac{\Delta}{2}\right) \Gamma\left(\Delta_b-\frac{\delta}{2}\right) M(\delta) .
    \label{eq:A3}
 \ea
So we have obtained a simple formula for the OPE coefficients in this special case where $\ell = 0$ and the Mellin amplitude is independent of $\gamma$  
\ba \label{eq:SpinZeroOPEExtraction}
 c^{34}_{\Delta, 0} &=& \frac{-\pi^{2h}}{4 N^{f12}_{\Delta, 0}} \frac{\Gamma^2(h-\frac{\Delta}{2})\Gamma^2(\frac{\Delta}{2})}{\Gamma(h)\Gamma(2h-\Delta)\Gamma(\Delta)}
     \label{eq:OPEcoeffs}\\
& & \times     \int \frac{d \delta}{2\pi i}  \Gamma\left( \frac{\delta}{2}-\frac{\Delta}{2}\right)  \Gamma\left(\frac{\delta}{2} -h+\frac{\Delta}{2}\right) \Gamma\left(\Delta_a-\frac{\delta}{2}\right)  \Gamma\left(\Delta_b-\frac{\delta}{2}\right) M(\delta) .  \nn
\ea
One can obtain the normalization $N^{f12}_{\Delta, 0}$ from equation (\ref{eq:SpinZeroNorm}), which can be computed term-by-term in perturbation theory if such a series is available.  We give the general formula for these OPE coefficients with arbitrary $\ell$ and a $\gamma$-dependent Mellin amplitude in appendix \ref{app:General4ptConglomeration}.

\subsubsection{Conformal Block Coefficients}

Let us discuss an application of this formalism to the extraction of conformal block coefficients and anomalous dimensions of double-trace operators at leading order in perturbation theory for a simple AdS theory.  Such examples were studied in \cite{JP,Katz,Fitzpatrick:2011hh}, where different methods were used. While \cite{Katz,Fitzpatrick:2011hh} found a fairly simple method for extracting anomalous dimensions, the calculation of conformal block coefficients remained cumbersome, to say the least.  We will begin with the decomposition of the four-point function into double-trace conformal blocks at leading order in $1/N$, although we will see later that this decomposition also has a non-perturbative interpretation:
\ba
\CA_4 = \sum_n P_1(n) B_{\Delta_n}(x_i) + \frac{1}{2} P_0 (n) \gamma(n) \frac{\partial}{\partial n}  B_{\Delta_n}(x_i) ,
\label{eq:fourpointconf}
\ea
where $P_0(n)= \bar{c}^{12}_{n,0} \bar{c}^{34}_{n,0}$ and $P_1(n) = c^{12}_{n, 0} \delta c^{34}_{n,0}$ are, respectively,  the infinite $N$ and correction terms to the double-trace conformal block coefficients. The partial derivative with respect to $n$ brings down logarithms, since $B_{\Delta_n}$ in position space contains terms with $x_i$'s to the $n$-th power.  
When we conglomerate the four-point function with $\Delta = \Delta_n$, we pick up the contribution from a specific double-trace operator.   
We will see later that this form is appropriate for the study not just of the leading order in perturbation theory, but furthermore for non-perturbative corrections to a large, important class of conformal block contributions that we will refer to as ``cuts through the edge of a diagram''.  Thus, the reader should keep in mind that although the specific examples we will compute in this section are perturbative, the general formulae we obtain will be applicable for gaining non-perturbative information about the CFT.  

The extraction of OPE coefficients in the presence of anomalous dimensions is a bit subtle.  Let us therefore begin with
a conceptually simpler case, where $\Delta_a \equiv \frac{\Delta_1 + \Delta_2}{2}$ and $\Delta_b \equiv \frac{\Delta_3 + \Delta_4}{2}$ are unrelated to each other.  Then, $\CO_1 \CO_2$ and $\CO_3 \CO_4$ do not have any of the same double-trace operators in their leading order OPE, so no anomalous dimensions will appear in the four-point function at this order.  From equations (\ref{eq:OPEcoeffs}) and (\ref{eq:c2overc3}), we find the following expression for $P_1$:
\ba
P_1(n) &=& \frac{-1}{4\Gamma(0)} \overbrace{\frac{\Gamma^4(\Delta_a+n)}{\Gamma(2(\Delta_a+n))\Gamma(2\Delta_a+2n-h)} }^{G(\Delta_a+n)} \\
  && \times \pi^2 \int \frac{d\delta}{2\pi i} M(\delta) \frac{
 \Gamma\left(\frac{\delta}{2}-\Delta_a-n\right)  \csc \left( \pi \left( \frac{\delta}{2} -\Delta_a \right)\right)\Gamma\left( \frac{\delta}{2}+\Delta_a + n -h\right)}
 {\Gamma\left(\frac{\delta}{2} - \Delta_a +1 \right) \Gamma\left( \frac{\delta}{2} - \Delta_b +1 \right)\sin\left( \pi \left( \frac{\delta}{2} -\Delta_b \right) \right)}. \nn
 \label{eq:singp1form}
\ea
Because of the singular $\Gamma(0)^{-1}$ prefactor, when we perform the $\delta$ integral by contour integration, we can discard any residues that are non-singular when $n$  is an integer.  This is a great simplification, because the only such contributions are those poles at
\ba
\frac{\delta}{2} = \Delta_a+m, \ \ \ \ \ \ m=0, \dots, n,
\ea
where there is a pole from both the $\csc$ term and the first $\Gamma$ function in the integrand.  These residues are in one-to-one correspondence with the residues of the following equivalent, but simpler, integral, where the $\Gamma^{-1}(0)$ has been cancelled:
\ba
P_1(n) &=& 
 \frac{-1}{4} \frac{G(\Delta_a+n)}{\sin \left( \pi \left( \Delta_a - \Delta_b \right) \right)}  \pi \int \frac{d\delta}{2\pi i} M(\delta) \frac{
 \Gamma\left(\frac{\delta}{2}-\Delta_a-n\right)  \Gamma\left( \frac{\delta}{2}+\Delta_a + n -h\right)}
 {\Gamma\left(\frac{\delta}{2} - \Delta_a +1 \right) \Gamma\left( \frac{\delta}{2} - \Delta_b +1 \right)} 
 \label{eq:p1nformula}
\ea
For any specific Mellin amplitude of the form $M(\delta)$, this formula for the OPE coefficients is relatively simple to use, since now for any $n$ it is just a finite sum of non-singular residues.  

\subsubsection{Anomalous Dimensions and the Derivative Relation}

We will now generalize the results in the previous section to include cases with anomalous dimensions.  This will requires addressing the subtlety mentioned above.  To see the issue explicitly, recall that the perturbative three-point function for two single-trace operators $\CO_1, \CO_2$ of dimension $\Delta_a$ with a double-trace operator of dimensions $\Delta+ \gamma(\Delta)$ takes the form
\ba
\CA_3 &=& \frac{\bar c_{n,0}^{12}+ \delta c_{n,0}^{12}}{P_{12}^{\Delta_a} } u ^{\frac{\Delta +\gamma(\Delta)}{2}} \\
&=& \frac{1}{P_{12}^{\Delta_a}} \left( u^{\frac{\Delta}{2}} \left( \bar c_{n,0}^{12} + \delta c_{n,0}^{12}  + \bar c_{n,0}^{12}  \frac{1}{2} \gamma(\Delta) \log u + \dots \right) \right),   \nn
\label{eq:threepointgamma}
\ea
where $u \equiv  \frac{P_{12}}{P_{01} P_{02}} $.  
The problem is that the position-dependence of a three-point function from a Mellin amplitude is completely fixed, and cannot contain any logarithms.  
 This is because at any finite order in
perturbation theory, the anomalous dimensions naively appear to break the conformal invariance, and it is only the resummation of all order of the logarithms
that reproduces a conformally invariant correlation function with the new, shifted dimensions of operators.   

However, we are actually in a position to get around this difficulty very easily with the use of the results we have just obtained above.  While there were no anomalous dimensions
for $\Delta_a$ and $\Delta_b$ unrelated, in the limit of $\Delta_b \rightarrow \Delta_a$ we should be able to see logarithms reappear.  The important physical point to note however is that now there is no difference between the double-trace operator with dimension $2\Delta_a+2n$ and the one with dimension $2 \Delta_b +2n$, so the physical three-point function will be the sum of both of these $P_1(n)$'s.  As one can see from the $\csc ( \pi (\Delta_a- \Delta_b))$ prefactor in eq. (\ref{eq:p1nformula}), each of these $P_1(n)$'s is singular in this limit. But, this singularity exactly cancels in their sum, and the subleading (finite) term in the three-point function contains a logarithm!  Evaluating this explicitly, we find the sum of the two three-point functions as $\Delta_b \rightarrow \Delta_a$ is
\ba
&&  \frac{1}{P_{12}^{\Delta_a}}  \lim_{\Delta_b \rightarrow \Delta_a} \left(p_1(n) u^{\Delta_a+n} + (\Delta_b \leftrightarrow \Delta_a) \right) \nn\\
 && = -\frac{1}{4P_{12}^{\Delta_a}} \lim_{\Delta_b \rightarrow \Delta_a } \left(\left(  \frac{\partial}{\partial \Delta_a}-\frac{\partial}{\partial \Delta_b} \right)u^{\Delta_a+n} G(\Delta_a+n) \int d\delta M(\delta) \frac{
 \Gamma\left(\frac{\delta}{2}-\Delta_a-n\right)  \Gamma\left( \frac{\delta}{2}+\Delta_a + n -h\right)}
 {\Gamma\left(\frac{\delta}{2} - \Delta_a +1 \right) \Gamma\left( \frac{\delta}{2} - \Delta_b +1 \right)} \right) \nn\\
 && = -\frac{1}{4 P_{12}^{\Delta_a}}  \frac{\partial}{\partial n}  \left( u^{\Delta_a+n} G(\Delta_a+n) \int d\delta M(\delta) \frac{
 \Gamma\left(\frac{\delta}{2}-\Delta_a-n\right)  \Gamma\left( \frac{\delta}{2}+\Delta_a + n -h\right)}
 {\Gamma\left(\frac{\delta}{2} - \Delta_a +1 \right) \Gamma\left( \frac{\delta}{2} - \Delta_b +1 \right)} \right).
\ea
Comparing this to the expected form of the three-point function in eq. (\ref{eq:threepointgamma}), we see that we have derived a simple formula
for the anomalous dimensions $\gamma(n)$ and OPE coefficients $p_1(n)$ when $\Delta_a = \Delta_b$:
\ba
P_0(n) \gamma(n) &=& -\frac{1}{2} G(\Delta_a+n)\int d\delta M(\delta) \frac{
 \Gamma\left(\frac{\delta}{2}-\Delta_a-n\right)  \Gamma\left( \frac{\delta}{2}+\Delta_a + n -h\right)}
 {\Gamma\left(\frac{\delta}{2} - \Delta_a +1 \right) \Gamma\left( \frac{\delta}{2} - \Delta_b +1 \right)}, \nn\\
P_1(n) &=& \frac{1}{2} \frac{\partial}{\partial n} P_0(n) \gamma(n). 
\ea

This proves the relation between OPE coefficients and anomalous dimensions that was found empirically in \cite{JP}.  While we have focused in this section on cases with only spin-0 conformal blocks, the proof in fact generalizes straightforwardly to any spin.  The reason is that 
this result depended only on two properties of our expression for the OPE coefficients: first, that only the singular residues in eq. (\ref{eq:singp1form}) survive the $\Gamma^{-1}(0)$ prefactor, and second, that these residues depend only on $\Delta_a$ through the combination 
$\Delta_a+n$, except for the factor $\Gamma\left( \frac{\delta}{2} -\Delta_a+1\right)\Gamma\left( \frac{\delta}{2} -\Delta_b+1\right)$ that is symmetric in $(\Delta_a \leftrightarrow \Delta_b)$. This allowed us to exchange a derivative $\Delta_a$ for one in $n$, since $\frac{\partial}{\partial \Delta_a}$ derivatives acting on this symmetric factor are cancelled by the $\frac{\partial}{\partial \Delta_b}$ derivative.

\subsubsection{Example Computations}
\label{sec:ExampleConglomerationCalcs}

Let us now apply this formula to some concrete examples.  The simplest possible AdS interaction that affects only scalar conformal blocks is
a $\lambda \phi^4$ interaction, which corresponds to a Mellin amplitude that is just a constant.  
Let us now apply this formula to some concrete examples.  The simplest possible AdS interaction that affects only scalar conformal blocks is
a $\lambda \phi^4$ interaction, which corresponds to a Mellin amplitude that is just a constant.  Then, the anomalous dimension is simply 
\ba
\gamma(n) &\propto& G(\Delta_a + n ) \sum_{m=0}^n \frac{(-1)^{m+n}}{(n-m)!(m!)^2} \Gamma(2\Delta_a + n+m-h) \nn\\
&\propto& \frac{1}{\left( \bar c^{12}_{n,0} \right)^2 } \frac{(h)_n (2\Delta_a+n-2h+1)_n (2\Delta+2n)_{-h}}{(\Delta_a+n)_{1-h}^2 (2\Delta_a + n-h)_n},
\ea
where we have used the expression for the infinite $N$ OPE coefficients from eq. (\ref{eq:scalarOPEinfN}) and dropped an overall $n$-independent prefactor.  This quantity was computed using alternate methods
in \cite{Katz}, whose results can be seen to agree with that above. 

Next, let us turn to $\frac{\lambda}{4} \phi^2 \chi^2$, which has different fields on the left and right and is one of the contact interactions in figure \ref{fig:WittenDiagramtoBlocks}.  We will compute the OPE coefficients $\delta c_{2\Delta_\chi + 2n, 0}^{\phi \phi}$ at first order in $\lambda$; these will be useful ingredients when we study an example of the optical theorem in section \ref{sec:OneLoopExample}.  In this case, we apply equation (\ref{eq:SpinZeroOPEExtraction}) to the trivial Mellin amplitude $M = \lambda_4\equiv \lambda \frac{\pi^h}{2} \Gamma( \Delta_\Sigma- h) \prod_{i=1}^4 \frac{\CC_{\Delta_i}}{\Gamma(\Delta_i)}$.  We can normalize the wavefunction $f^{\phi \phi}_{\Delta, 0}$ by using equation (\ref{eq:SpinZeroNorm}) with the mean field theory $\bar c^{\chi \chi}_{2 \Delta_\chi + 2n,0}$.  For the normalization of the wavefunction we find
\be
N^{f\chi \chi}_{\Delta, \ell} = 
\frac{\pi^{2h} \Gamma(0)\Gamma \left(h- \frac{\Delta}{2} \right)^2 \Gamma(\Delta -h) }{\Gamma(h) \Gamma \left(\frac{\Delta}{2} \right)^2  \Gamma(2h-\Delta)}
\times \frac{ \CC_{\Delta_\chi}(\Delta_\chi)_n \left(1 + \Delta_\chi-h \right)_n}{\sqrt{ n! 
   \left(h \right)_n  (2\Delta_\chi - 2h +1+n)_n  \left(2 \Delta_\chi - h+ n\right)_n} }  
\ee
where $\Delta = 2 \Delta_\chi + 2n$ corresponds to the dimension of $[\CO_\chi \CO_\chi]_{n, 0}$ to first order in perturbation theory.  To apply equation (\ref{eq:SpinZeroOPEExtraction}) we need only integrate using Barnes' Lemma, giving 
\ba \label{eq:TreeLevelc3}
\delta c^{\phi \phi}_{\Delta, 0} &=& 
\lambda_4 \frac{(-1)^{n+1} \Gamma(-n+\Delta_\phi-\Delta_\chi) \Gamma^4(\Delta_\chi+n) \Gamma(\Delta_\chi+\Delta_\phi +n-h) \Gamma(2\Delta_\chi + n -h)  }{2n! \CC_{\Delta_\chi} \Gamma(2\Delta_\chi + 2n) \Gamma(\Delta_\phi + \Delta_\chi -h) \Gamma(2\Delta_\chi + 2n-h) (\Delta_\chi)_n (\Delta_\chi-h+1)_n} \nn\\
&& \times \sqrt{n! (h)_n (2\Delta_\chi+n-2h+1)_n (2\Delta_\chi +n-h)_n }
\ea
We can use this coefficient and the equivalent one with $\phi \to \psi$, to compute one-loop conformal block coefficients, as pictured in figure \ref{fig:WittenDiagramtoBlocks}.  This will be useful for verifying the unitarity relation in the flat space limit when we come to section \ref{sec:OneLoopExample}.

\section{S-Matrix Unitarity from CFT Unitarity}
\label{sec:SMatrixUnitarity}

\begin{figure}[t!]
\begin{center}
\includegraphics[width=0.95\textwidth]{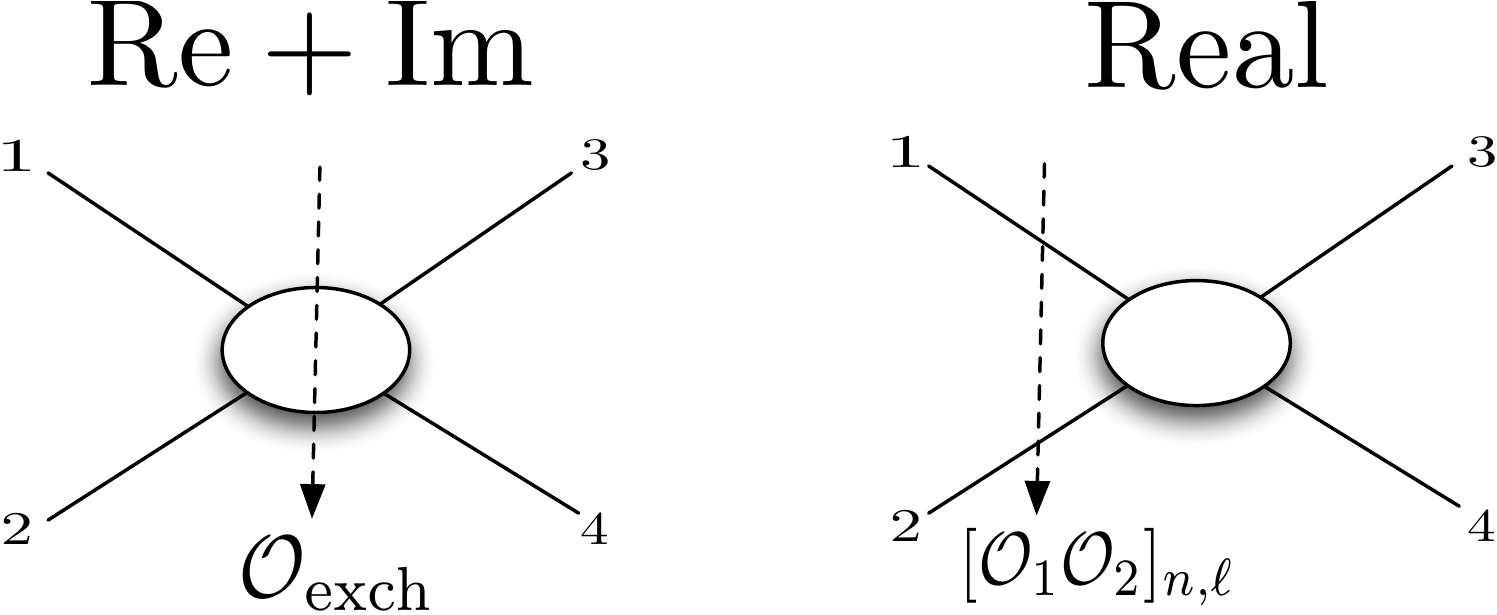}
\caption{ This figure depicts how the `edge cuts', which correspond to the terms in the conformal block decomposition involving free propagation, only contribute to the real part of the bulk S-Matrix, while other operator exchanges contribute to both the real and the imaginary pieces of the S-Matrix in the flat space limit of AdS/CFT.
\label{fig:EdgevsCenterCuts}  }
\end{center}
\end{figure}

In this section we will derive the optical theorem 
\be \label{eq:OpticalTheoremBody}
-i \left( T - T^\dag \right) = T^\dag T
\ee
for 2-to-2 scattering of massless scalars by analyzing the conformal block decomposition in the flat spacetime limit of the dual AdS theory.  The derivation will occur in several steps.  First, in section \ref{sec:FlatLimitofBlocks} we review our recent result from \cite{Fitzpatrick:2011hu}, where we showed that in the flat space limit of AdS/CFT, conformal blocks correspond to delta functions in the center of mass energy with a definite angular momentum $\ell$.  This means that 
\be
\sum_{\Delta, \ell} P_{\Delta, \ell} B_{\Delta}^\ell(\delta_{ij}) \ \  \stackrel{R \to \infty}{\longrightarrow} \ \ \mathcal{S} \propto P_{\sqrt{s}, \ell} C_{\ell}^{(h-1)} (\cos \theta)
\ee
We will carefully compute the phase in the normalization of the blocks in order to precisely identify their imaginary parts.  Then in section \ref{sec:ImaginaryPartofSMatrix} we will compute the left-hand side of equation (\ref{eq:OpticalTheoremBody}) in terms of the conformal block decomposition.  Terms in the conformal block decomposition that we call `edge cuts' only contribute to the real part of the S-Matrix\footnote{We will refer to the left-hand side of equation (\ref{eq:OpticalTheoremBody}) as the imaginary part of the S-Matrix, although in fact it can be complex if the in and out states are distinct.};  these edge cuts are pictured in figure \ref{fig:EdgevsCenterCuts} and also shown in a perturbative example in figure \ref{fig:WittenDiagramtoBlocks}.  The conformal block coefficients are simply products of OPE coefficients, so that schematically
\be
P_{\sqrt s, \ell} = \left(\bar c^L_{\sqrt s, \ell} + \delta c^L_{\sqrt s, \ell} \right) \left( \bar c^R_{\sqrt s, \ell} + \delta c^R_{\sqrt s, \ell} \right)
\ee
In this expression, the edge cuts are simply the terms that involve the mean field theory OPE coefficients $\bar c_{\sqrt s, \ell}$.  To prove that the edge cuts drop out of the optical theorem we will make essential use of the relations we derived in section \ref{sec:FurtherApplications} for the double-trace conformal block coefficients and their anomalous dimensions.  

The central cuts pictured in figure \ref{fig:EdgevsCenterCuts} do contribute to the imaginary part of the S-Matrix, so it remains to show that these are equal to the right hand side of the optical theorem.  The central cuts are exactly the conformal block coefficients given by the product of interacting OPE coefficients, so we have that
\be
\left. -i \left( T - T^\dag \right) \right|_{\sqrt s, \ell} = \sum_{\stackrel{\CO_{\Delta, \ell}}{\Delta \approx R \sqrt s}} \delta c^{12}_{\Delta, \ell} \delta c^{34}_{\Delta, \ell}
\ee
But the right hand side is already in a form that can be interpreted as the right hand side of optical theorem, $T^\dag T$.  It only remains to argue that the OPE coefficients are proportional to scattering amplitudes, and that the sum over exchanged operators corresponds to a phase space integral over multi-particle states.  The OPE coefficients $\delta c_{\Delta, \ell}$ can be computed by conglomerating $k+2$-point correlators into 3-point correlators involving $k$-trace operators, and these $k$-trace operators provide a basis for scattering states in the flat space limit \cite{Fitzpatrick:2011jn}.  We will show in section \ref{sec:PhaseSpace} that the sum over $k$-trace operators becomes a phase space integral over $k$-particle states in the flat space limit of AdS \cite{Fitzpatrick:2011jn}, as depicted in figure \ref{fig:OperatorSumtoPhaseSpace}, so that
\be
\sum_{\stackrel{\CO_{\Delta, \ell}}{\Delta \approx R \sqrt s}} \delta c^{12}_{\Delta, \ell} \delta c^{34}_{\Delta, \ell}   \ \ 
\stackrel{R \to \infty}{\longrightarrow} \ \ 
\sum_{k=1}^\infty \int \prod_{i=1}^k \frac{d^d q_i}{(2 \pi)^d 2 E_i} \delta^{d+1} \left(p_1 + p_2 - \Sigma_i q_i \right) \mathcal{M}_{12 \to k} \mathcal{M}^*_{34 \to k}
\ee
This will complete the derivation of the optical theorem.  Finally, in section \ref{sec:OneLoopExample} we show how this logic applies in a complete one-loop example.

\subsection{Conformal Blocks in the Flat Space Limit of AdS/CFT}
\label{sec:FlatLimitofBlocks}

The conformal block decomposition will be crucial to our proof of the unitarity of the holographic S-matrix.  As we discussed in the introduction, a conformal block corresponds to the exchange of a particular state in the CFT, and in perturbation theory conformal blocks sum up to give the various different cuts of Feynman diagrams.  So, we will first review the flat-space limit of conformal blocks from \cite{Fitzpatrick:2011hu}, paying close attention to the phase in the normalization of the result.  It was shown there that the flat-space limit of a block is proportional to a $\delta$ function in the center of mass energy.  We explicitly performed the transformation between the Mellin amplitude and the S-Matrix \cite{Penedones:2010ue, Fitzpatrick:2011hu} 
\ba
T(s_{ij}) &=&  \lim_{R \to \infty} \frac{1}{\CN} \int_{-i\infty}^{i\infty} d \alpha \ \!  e^{\alpha}  \alpha^{h - \Delta_\Sigma}  
M \left( \delta_{ij} = -\frac{R^2 s_{ij}}{4 \alpha} \right), \nn\\
\CN &=& \frac{\pi^h R^{\frac{n(1-d)}{2} + d +1} }{2} \prod_{i=1}^n \frac{\CC_{\Delta_i} }{\Gamma(\Delta_i )} \ \ \stackrel{n=4} {\longrightarrow} \ \ 2^{-5} \pi^{-3h}R^{3-2h} \prod_{i=1}^4 \frac{1}{\Gamma(\Delta_i +1-h)} ,
\label{eq:flatspacetransform}
\ea
where $\Delta_\Sigma \equiv \frac{1}{2} \sum_i \Delta_i$ for half the sum of the external dimensions. 
Alternately, one can argue based on general principles as follows.  Conformal blocks are just the contribution to
the four-point function (or $n$-point functions, more generally) from complete irreducible representations of the conformal
group.  Consider a conformal block whose primary has dimension $\Delta$.  The primaries are the lowest weight states of
the representation, i.e. those annihilated by the generators $K^i$ of special conformal transformations.  
There is a one-to-one mapping of AdS states to CFT states, as well as of generators of the conformal algebra to generators
of AdS isometries, which in the flat space limit is just the Poincar\'e algebra. Furthermore, special conformal generators
in the flat-space limit become the momentum operators \cite{Katz, Fitzpatrick:2011jn}.  
Thus, all states in the conformal block map onto a complete irreducible representation of the Poincar\'e group that
has center-of-mass energy $\sqrt{s}= \Delta/R$.  Consequently, in the flat-space limit, the conformal block
can contribute only at this value of $s$.  Its  angular dependence is further constrained by symmetry to be the appropriate
polynomials in $\cos \theta$, which are Legendre polynomials in $d=3$ and Gegenbauer polynomials $C_\ell^{(h-1)}(\cos\theta)$ more generally.  

To read off the normalization, it is simplest to use the inverse of eq. (\ref{eq:flatspacetransform}), because integrating over delta functions is very easy.  So we compute
\be
M(\delta_{ij}) \stackrel{\delta_{ij} \gg 1}{= } \CN \int_0^\infty \frac{d\beta}{\beta} e^{-\beta} \beta^{\Delta_\Sigma-h} T\left(s_{ij} = -4 \beta \frac{\delta_{ij}}{R^2}\right) 
\ee
and input the form for $T(s_{ij})$  required by symmetry:
\ba
T(s_{ij}) &=& A_{\Delta, \ell} \delta(s - \Delta^2/R^2) C_\ell^{(h-1)}(\cos\theta) .
\ea
The integration over $\beta$ is then trivial to perform:
\ba
\CN^{-1} M(\delta_{ij}) &\stackrel{\delta_{ij} \gg 1} {=} &\left[  A_{\Delta, \ell} \left(\frac{\Delta^2}{4}\right)^{\Delta_\Sigma -h -1} \frac{R^2}{4} \right] (-\delta_{12})^{h-\Delta_\Sigma} e^{\frac{\Delta^2}{4 \delta_{12}}} C_\ell^{(h-1)}(\cos \theta) %\nn\\
\label{eq:inverseConfBlock}
\ea  
This result should be compared to the large $\delta_{ij}$ limit of the conformal blocks themselves.
They are fixed up to an overall normalization by conformal invariance, and we will normalize them in accordance with our definitions from previous sections, so that
\ba
B_\tau^\ell(\delta_{ij}) = \frac{ e^{ \pi i (h- \tau+1)} \left( e^{i \pi ( \delta + \tau - 2h)} - 1 \right)}{2\pi i} \frac{\Gamma(\Delta)\Gamma(\Delta-h+1)}{\Gamma^4\left( \frac{\Delta}{2} \right)} \frac{ \Gamma\left( \frac{ \tau - \delta}{2} \right) \Gamma\left( \frac{2h - \tau -2\ell - \delta}{2} \right) } { \Gamma \left( \Delta_a - \frac{\delta}{2} \right) \Gamma\left(\Delta_b - \frac{\delta}{2} \right) } P_{\ell, \tau}(\delta_{ij}),
\ea
where we define $\Delta_1=\Delta_2=\Delta_a$ and  $\Delta_3=\Delta_4=\Delta_b$.  Here, $\tau=\Delta-\ell$ is the twist of the conformal block and $P_{\ell, \tau}(\delta_{ij})$ is a Mack polynomial \cite{Mack, Fitzpatrick:2011hu}. For $\ell =0$, it is just $P_{0,\Delta}=1$.  We will need only the large $\delta_{ij}$ limit of the block, with $\delta_{ij} \propto s_{ij}$:
\ba
P_{\ell, \tau}(\delta_{ij}) & \stackrel{\delta_{ij} \gg 1}{=}& g_{\ell, \tau} (-\delta_{12})^\ell C_{\ell}^{(h-1)} (\cos \theta) ,
\ea
where the proportionality constant $g_{\ell, \tau}$ is real and $g_{0,\Delta}=1$. Expanding $B_\tau^\ell(\delta_{ij})$ at large $\delta_{ij}$ and $\Delta$, we obtain the approximation
\ba
\CN^{-1} B_\tau^\ell (\delta_{ij})& \stackrel{\delta_{ij} \gg 1}{=} &\pi^{3h-1}  \frac{g_{\ell, \tau}\Delta^{2-h}(-1)^\ell}{\prod_{i=1}^4 \Gamma(\Delta_i+1-h)} 2^{3+2\Delta} R^{2h-3} (-\delta_{12})^{h-\Delta_\Sigma} e^{\frac{ \Delta^2}{4 \delta_{12}} } C_\ell^{(h-1)} (\cos \theta)  \nn\\
&& \times \left(i - \cot \left( \frac{\pi}{2} (2 \delta_{12} -2 \Delta_a + \tau) \right) \right)\sin (\pi \delta_{12}) \sin( \pi (\delta_{12}-\Delta_a + \Delta_b) ).
\ea
Because of the $\sin$ and $\cot$ factors, this does not strictly speaking have a well-defined large $\delta_{12}$ limit.  However, if we smooth over an $\CO(1)$ region of $\delta_{12}$, the last line averages out to
\be
\frac{-i \cos (\pi (\Delta_a - \Delta_b)) + \sin (\pi (\Delta_a + \Delta_b-\tau))}{2} .
\label{eq:confblockflatcoeff}
\ee
As an aside that will be relevant shortly, note that for the double-trace operators $\tau= 2\Delta_a+2n+\ell$ and $\tau=2\Delta_b+2n+\ell$,
this simplifies further to
\be \label{eq:EvenOddness}
-i e^{(-1)^{\ell+1} i \pi (\Delta_a -\Delta_b)}  \ \ \ \ \mathrm{and} \ \ \ \ -i e^{ (-1)^\ell i \pi (\Delta_a -\Delta_b)},
\ee
respectively. Thus we have obtained the overall normalization coefficient for the flat space limit of the conformal blocks 
\ba
\left[  A_{\Delta, \ell} \left(\frac{\Delta^2}{4}\right)^{\Delta_\Sigma -h -1}  \right] &=& 
  \frac{\pi^{3h-1}  2^{5+2\Delta} R^{2h-5} \Delta^{2-h} }{ \prod_{i=1}^4 \Gamma(\Delta_i+1-h)} 
  \\
  & & \times \left( \frac{-i \cos (\pi (\Delta_a - \Delta_b)) + \sin (\pi (\Delta_a + \Delta_b-\tau))}{2} \right)  \nn
  \label{eq:Adelta}
\ea
Note that both terms inside the parentheses appear to be even functions when we switch $\Delta_a \leftrightarrow \Delta_b$. However, in the special case where the conformal block is a double-trace operator, it is important to note that when we make this switch we must take $\tau$ from $2 \Delta_a + 2n$ to $2 \Delta_b + 2n$, as was already made manifest in equation (\ref{eq:EvenOddness}).  Thus we observe the crucial fact that the imaginary part of the coefficient for double-trace operators remains even under $\Delta_a \leftrightarrow \Delta_b$, while the real part is odd.

\subsection{The Imaginary Part of the S-Matrix}
\label{sec:ImaginaryPartofSMatrix}

Let us use the tools we have developed to derive the optical theorem. 
As we saw in the previous section, when we take the flat space limit of the conformal block decomposition of the 4-pt correlator we find
\ba \label{eq:FSLofSumOverBlocks}
T_{12 \to 34}(s_{ij}) &=& \sum_{\Delta, \ell}  P_{\Delta, \ell} \left[ A_{\Delta, \ell} \delta \left(s - \Delta^2/R^2 \right) C_\ell^{(h-1)}(\cos\theta) \right]
\ea
The normalization factor $A_{\Delta, \ell}$ was obtained in equation (\ref{eq:Adelta}), and $P_{\Delta, \ell}$ is the full conformal block coefficient.  

In general, all operators $\CO_{\Delta, \ell}$ with the charge of $\CO_1 \CO_2$ can contribute to this sum, and so for each dimension $\Delta$ we can  write
\be \label{eq:GeneralBlockCoeff}
P_{\Delta, \ell} = c^{12}_{\Delta, \ell} {c^{34}_{\Delta, \ell}}^{\! \! \! *} 
\ee
We have expressed the conformal block coefficients in terms  of CFT 3-pt functions by using the OPE, as discussed in section \ref{sec:UsingConglomeration}.  Now we would like to isolate the double-trace operators $[\CO_1 \CO_2]_{n, \ell}$ and $[\CO_3 \CO_4]_{n, \ell}$.  These operators play a special role because they are present even when the bulk theory is free, and so they are responsible for the `$1$' when we write $S = 1 + iT$.  In the special case of these operators, we can write
\be
c_{n, \ell}^{12} = { \bar c_{n, \ell}^{12}} + \delta c_{n, \ell}^{12}
\ee
where $ \bar c_{n, \ell}^{12}$ is the 3-pt function coefficient corresponding to a free bulk theory and $\delta c_{n, \ell}^{12}$ is the change in this coefficient due to interactions.  We will use a similar notation for $c_{n, \ell}^{34}$, although note that $[\CO_1 \CO_2]_{n, \ell}$ and $[\CO_3 \CO_4]_{n, \ell}$ will, in general, have different dimensions and so give distinct conributions.  Now we can write the conformal block coefficients for these operators as
\be
P_{n, \ell} = { \bar c_{n, \ell}^{12}} { \bar c_{n, \ell}^{34}} + \left( { \bar c_{n, \ell}^{12}} \delta c_{n, \ell}^{34} + \delta c_{n, \ell}^{12} \bar c_{n, \ell}^{34} \right) + \delta c_{n, \ell}^{12} \delta c_{n, \ell}^{34} 
\ee
If the operators $\CO_1$ and $\CO_2$ are the same as $\CO_3$ and $\CO_4$, then the first term corresponds exactly to the `$1$' part of the S-Matrix, and otherwise it is absent.  The final term comes purely from interactions, and actually combines two different pieces, one involving the exchange of $[\CO_1 \CO_2]_{n, \ell}$ and the other involving the exchange of $[\CO_3 \CO_4]_{n, \ell}$.  However, the terms in parentheses are precisely the edge cuts pictured in figure \ref{fig:EdgevsCenterCuts}.  They combine free propagation on one side with interactions on the other, and are associated with poles in the $\Gamma$ functions from the Mellin integrand rather than poles in the Mellin amplitude itself.\footnote{
This definition of ``edge cuts'' as any term with a $\bar{c}$ factor should be intuitively reasonable, but we can also more rigorously connect it to bulk diagrammatics.  For any diagram, we can formally label all internal field lines by $\phi_i^{(I)}$'s, which are distinct from the external fields $\phi_i^{(E)}$.  This is just a relabeling and does not change the Mellin amplitude itself.  However, it is now manifest that ``cuts through the middle of the diagram'' are any conformal blocks for operators made of internal fields $\phi_i^{(I)}$, and ``cuts through the edge of the diagram'' are any conformal blocks for operators made of external fields $\phi_i^{(E)}$.  Since $\phi_i^{(E)}$'s never appear as internal lines, such a diagram is not sensitive to any lower order corrections to their conformal blocks, and must take the form of a leading correction, i.e.
\ba
\CA &\supset& \bar{c}^{12}_{n,\ell} \delta c^{34}_{n,\ell} B_{2\Delta_a+2n} + \delta c^{12}_{n,\ell} \bar{c}^{34}_{n,\ell} B_{2\Delta_b+2n},
\ea
plus a possible $\frac{1}{2} (\bar{c}^{12})^2 \gamma_{12} \frac{\partial}{\partial n}B_{2\Delta_a+2n}$ term if $[\CO_1 \CO_2]_{n, \ell} = [\CO_3 \CO_4]_{n', \ell}$.  
This is a long-winded way of saying that in this labeling, it is manifest that ``edge cuts'' are exactly equivalent to terms that contain $\bar{c}^{12}$ or $\bar{c}^{34}$.  However, $\bar{c}$'s are exactly identified as the parts of the OPE coefficients that are zero-th order in any bulk couplings, and this characterization of them is completely unaffected by our formal relabeling of the fields. This proves the claim.  
Furthermore, since $\phi_i^{(E)}$'s never appear as internal lines, there will be no poles corresponding to them in the Mellin amplitude itself -- all their poles appear solely in the $\Gamma$ functions in the definition of the Mellin integrand.  This indicates that the appropriate non-perturbative definition of edge cuts is contributions to correlators from the poles in these $\Gamma$ functions.  }

Let us show that these edge cuts drop out of the imaginary part of the S-Matrix.  Assume for simplicity that we are dealing with real scalar fields, so all bulk couplings are real; the generalization to complex couplings is straightforward.  Now the OPE coefficients are real, and $i(T^\dagger - T) = 2 \CI m (T)$.  Furthermore, in any physical process in the flat-space limit, there will be a finite resolution much greater than the AdS curvature scale, which we can incorporate through a resolution function $f_\epsilon(s,s_0)$ narrowly peaked on $s=s_0$. Taking the flat-space limit of the conformal block decomposition and integrating against this resolution, the edge cuts contribute as
\be
T_\epsilon 
\propto -i \sum_{n,\ell}   f_\epsilon((2\Delta_a+2n)^2/R^2, s_0)  \bar{c}^{12}_{n,\ell}  \delta c^{34}_{n,\ell} e^{- i \pi (-1)^\ell \Delta_{ab} } C^{(h-1)}_\ell(\cos \theta)  + 
(1,2, \Delta_a \leftrightarrow 3,4, \Delta_b) 
\ee
where $T_\epsilon$ indicates the finite resolution and we have used our computation of the normalization factors $A_{\Delta,\ell}$, defining $\Delta_{ab} \equiv  (\Delta_a-\Delta_b)$ for convenience. Only the phase of $A_{\Delta,\ell}$ is relevant here, so we have discarded a real overall coefficient. We are primarily interested in the imaginary piece of $T$ for unitarity, but it will be enlightening to keep track of both
its real and imaginary pieces.  We will now use our formula from equation (\ref{eq:p1nformula}) for the OPE coefficients.  This is specific to $\ell=0$, but the generalization of the following step to non-zero spins is straightforward:
\ba
T_{\epsilon}^{\ell=0} &\propto& \left[ \sum_n  f_\epsilon((2\Delta_a+2n)^2/R^2, s_0)   \frac{ -ie^{-i \pi \Delta_{ab}}G(\Delta_a+n)  }{\sin (\pi \Delta_{ab})} 
\int d \delta M(\delta) \frac{
 \Gamma\left(\frac{\delta}{2}-\Delta_a-n\right)  \Gamma\left( \frac{\delta}{2}+\Delta_a + n -h\right)}
 {\Gamma\left(\frac{\delta}{2} - \Delta_a +1 \right) \Gamma\left( \frac{\delta}{2} - \Delta_b +1 \right)} \right. \nn\\
&+& \left.  \sum_n  f_\epsilon((2\Delta_b+2n)^2/R^2, s_0)    \frac{-i e^{i \pi \Delta_{ab}}G(\Delta_b+n) }{\sin (-\pi \Delta_{ab})} 
\int d \delta M(\delta) \frac{
 \Gamma\left(\frac{\delta}{2}-\Delta_b-n\right)  \Gamma\left( \frac{\delta}{2}+\Delta_b + n -h\right)}
 {\Gamma\left(\frac{\delta}{2} - \Delta_a +1 \right) \Gamma\left( \frac{\delta}{2} - \Delta_b +1 \right)} \right] \nn\\
&\propto& G\left(\sqrt{\frac{s_0 R^2}{4} } \right) \int d\delta M(\delta) \frac{ \Gamma\left( \frac{\delta}{2} -  \sqrt{\frac{s_0 R^2}{4} }\right) \Gamma\left( \frac{\delta}{2} +  \sqrt{\frac{s_0 R^2}{4}} -h\right)}{\Gamma\left(\frac{\delta}{2} - \Delta_a +1 \right) \Gamma\left( \frac{\delta}{2} - \Delta_b +1 \right)}.
\ea
The two conformal blocks have combined to give the integral in brackets times a real coefficient!

The case where $\Delta_a=\Delta_b$ can be considered as a limiting case of $\Delta_a \ne \Delta_b$, and so is included in our proof above. 
However, one can still ask how the imaginary piece cancels technically in this case, since there are no longer two different conformal blocks
from the left and right side of the diagram to cancel against each other.  The resolution of this issue is that this is exactly the situation where
the derivative relation equation (\ref{eq:derivreln}) is satisfied.  Therefore, each conformal block contributes exactly as a total derivative:
\ba
\CA_4 &=& \sum_{n,\ell} \frac{\partial }{\partial n} \left( \left(\bar{c}^{12}_{n,\ell} \right)^2 \frac{1}{2}\gamma(n,\ell) B_{\Delta_n,\ell}(x_i) \right).
\ea
The imaginary piece of the conformal block coefficient eq.~(\ref{eq:confblockflatcoeff}) is smooth as an analytic function in $n$, 
and so when we take the flat-space limit the sum becomes an integral over a total derivative, and therefore it vanishes.  

Next we need to explain why the only contributions to equation (\ref{eq:GeneralBlockCoeff}) are from $k$-trace operators dual to states composed of stable bulk particles.  As we mentioned in the introduction, although unstable particles can be included in the perturbative cutting rules, the S-Matrix is only a well-defined unitary transformation between states made up of exactly stable particles.  Thus the operator $\CO_\chi$ dual to an unstable particle $\chi$ in AdS will not appear in the optical theorem, and it must make a vanishing contribution to the conformal block coefficients in the flat space limit of AdS.  This follows because $\CO_\chi$ will mix very quickly with the multi-trace operators into which it can decay; roughly speaking, if $\chi$ has a lifetime $\tau$, then it will only exist for a time of order $\tau/R$, so its contribution in the flat space limit will go to zero.  We saw this effect in the concrete example of a bulk $\mu \phi^2 \chi$ theory in \cite{Fitzpatrick:2011hu}, where we derived the Breit-Wigner resonance behavior from a re-summed Mellin amplitude.  In that case, for any finite AdS scale $R$ the single $\chi$ mode gave a finite contribution, but $\CO_\chi$ itself became negligible as $R \to \infty$, as it was replaced by the continuum of $2 \phi$ states.  Another familiar manifestation of this fact is that the delta function resonance from a stable particle is infinitely sharper, and therefore infinitely taller, than the smooth resonance from an unstable particle.  So while dropping the single mode corresponding to a stable particle would completely erase its delta function resonance, dropping an unstable particle mode has a negligible effect on the S-Matrix.  In summary: only operators dual to stable particles make an appearance in equation (\ref{eq:GeneralBlockCoeff}).

Finally, we will complete the argument by showing that the sum over $k$-trace operators in equations (\ref{eq:FSLofSumOverBlocks}) and (\ref{eq:GeneralBlockCoeff}) becomes a $d+1$ dimensional $k$-particle phase space integral. 

 \subsection{Sums Over Operators as Integrals Over Phase Space}
 \label{sec:PhaseSpace}
 
To complete our derivation of the optical theorem and of Cutkosky's `cutting rules', we need to show that the sum over the exchange of all $k$-trace operators turns into a phase space integral over $k$-particle states in the flat space limit of AdS/CFT, as pictured in figure \ref{fig:OperatorSumtoPhaseSpace}.  
 
We can understand this by noting that in the large $N$ limit, the space of states created by single-trace CFT operators is isomorphic to the Fock space of free particle states in AdS.  This follows from the fact that in AdS/CFT, the Hilbert spaces of the two theories are identical.   For example, if we quantize a free scalar field in AdS \cite{Banks:1998dd, Balasubramanian:1998de, Bena:1999jv, Fitzpatrick:2011jn}, we find
\be
\phi(t, \rho, \Omega) = \sum_{n,l,J} \phi_{nl}(t, \Omega,  \rho) a_{nlJ} + \phi_{nlJ}^*(t, \Omega,  \rho) a_{nlJ}^\dag
\ee
while the dual CFT operator can be quantized in terms of the same creation and annihilation operators, $a_{nlJ}^\dag$ and $a_{nlJ}$,  as
\be
\CO(t, \Omega) = \sum_{n,l,J} \frac{1}{N_{nlJ}^\CO}  \left( e^{i E_{n,l} t} Y_{lJ} (\Omega) a_{nlJ} +   e^{-i E_{n,l} t} Y_{lJ}^* (\Omega)a_{nlJ}^\dag \right)
\ee
We gave the explicit wavefunctions and normalizations in \cite{Fitzpatrick:2011jn}, but the crucial point is physical and independent of the details.  As is well known, particles in AdS behave as though they are in an IR-regulating cavity with a size set by the AdS length $R$.  Thus for finite $R$, the spectrum of $k$-trace states behaves like the discrete spectrum of $k$-particle states in a box of size $R$.  When we take the flat space limit $R \to \infty$, the discrete modes approach a continuum, and we recover the usual $k$-particle Lorentz invariant phase space when we sum over these modes.  An explicit analysis of the wave functions $\phi_{n, \ell}$ confirms this intuition \cite{Fitzpatrick:2011jn}, and the standard AdS quantization above reduces to the flat space quantization of a free field in spherical coordinates.  Since from AdS/CFT we know that the hilbert spaces of AdS particles and CFT states are identical, we can conclude that by summing over a complete set of particle states we are also summing over all possible CFT operators.

As a concrete example, in \cite{Katz} one of us considered this process in detail for the case of double trace operators and 2-particle states.  There it was shown that the state created by a double-trace operator $[\CO_1 \CO_2]_{n, \ell, J}$ can be expressed as
\be \label{eq:TwoParticlePhaseSpace}
|n, \ell, J \rangle = \frac{|2 p |^{\frac{d-2}{2}}}{(2 \pi)^d \sqrt{2 R E} } \int d \hat p Y_{\ell J}( \hat p) \int d^d q f(q)
|q+ p\rangle |q - p \rangle \ \ \ (n \gg 1)
\ee
where the state $|k \rangle$ is a one-particle state with momentum $k$, and the labels $J$ denote various angular momentum quantum numbers.  The important point is that for primary double-trace operators, the function $f(q)$ is fixed to be a Gaussian with width $\sqrt{E/R}$.  This means that in the flat space limit where $n = ER$ we obtain precisely the $\ell$th partial wave corresponding to a 2-particle state with center of mass energy $E$. 

As we have discussed above, the coefficient of the conformal blocks $B_{\Delta, \ell}$ at a dimension $\Delta$ can be computed by summing over the squares of appropriately normalized 3-pt functions 
\be 
\langle \CO_1 \CO_2 \CO_1 \CO_2 \rangle = \sum_{\ell, J} \left( c^{12}_{n\ell J} \right)^2 B_{n, \ell}   \ \ \ \ \mathrm{where} \ \ \ \  c^{12}_{n \ell J}T_{ \Delta_1, \Delta_2}^{n, \ell, J} = \left\langle \CO_1(P_1) \CO_2(P_2) \right| n, \ell, J \rangle
\ee
If we write the double-trace states $|n, \ell, J \rangle$ using equation (\ref{eq:TwoParticlePhaseSpace}), then in the flat space limit we can sum over the angular momentum quantum numbers $\ell, J$ while fixing the dimension $n \approx ER$ at the center of mass energy of the scattering process, as measured in AdS units.  This returns us from the spherical to the plane wave basis, giving
\ba \label{eq:2ParticlePhaseSpaceExample}
\sum_{\ell, J} \left( c^{12}_{ER, \ell, J} \right)^2 &\to& \left( \frac{|2 p |^{\frac{d-2}{2}}}{(2 \pi)^d \sqrt{2 R E} } \right)^2 \int d \hat p d \hat p'   \sum_{\ell, J} Y_{\ell J}( \hat p) Y_{\ell J}^*( \hat p')  c^{12}_{ER} \left( p, -p \right) {c^{12}_{ER}}^{\! \! \! *} \left(p', -p' \right) \nn  \\
&=& \int \frac{d^{d} \vec k_a}{(2 \pi)^d 2 |k_a|} \frac{d^{d} \vec k_b}{ (2 \pi)^d 2 |k_b|} \delta^{d+1} ( P_{\mathrm{CoM}} - k_a - k_b) \left| c^{12}_{ER}(k_a, k_b) \right|^2 
\ea
where the new 3-pt function coefficient $c^{12}_{ER}(k_a, k_b)$ in the plane wave basis can be re-interpreted as the square of the flat space scattering amplitude for $12 \to ab$, and $P_{\mathrm{CoM}}$ is the $d+1$ dimensional center of mass momentum.  To complete this re-interpretation, the external states must also be plane wave scattering states.  Plane waves with energy $\omega$ and velocity $\hat v$ are created by acting with  \cite{GGP, Fitzpatrick:2011hu} 
\be \label{eq:SingleParticleNorm}
|\omega, \hat v \rangle = \frac{2^{\Delta}\Gamma(\Delta)R^{\frac{d-3}{2}}}{(2\pi)^{h+1}\CC_\Delta (R \omega)^{\Delta-1}} \int_{ - \tau}^{\tau} dt
e^{i \omega t} \CO(t, -\hat v) |0 \rangle
\ee
Thus $c^{12}_{ER}(k_a, k_b)$ can only be interpreted as a scattering amplitude when the operators $\CO_1$ and $\CO_2$ are normalized and integrated in this way.
 
We should emphasize that the emergence of a $d+1$ dimensional phase space integral from the sum over operators is essentially kinematic.  It follows as a consequence of the structure of the conformal algebra as it reduces to the Poincar\'e algebra in the flat space limit of AdS/CFT, as we discussed in \cite{Fitzpatrick:2011jn}.  In particular, this means that if we instead study superconformal field theories, then we will instead be taking the flat space limit of the superconformal algebra.  Since the spacetime geometry follows from the algebra, in the future one should be able to obtain higher dimensional phase space integrals corresponding to decompactifying bulk dimensions from the flat space limit of superconformal theories.  

\subsection{A Complete One-Loop Example}
\label{sec:OneLoopExample}

\begin{figure}[t!]
\begin{center}
\includegraphics[width=0.99 \textwidth]{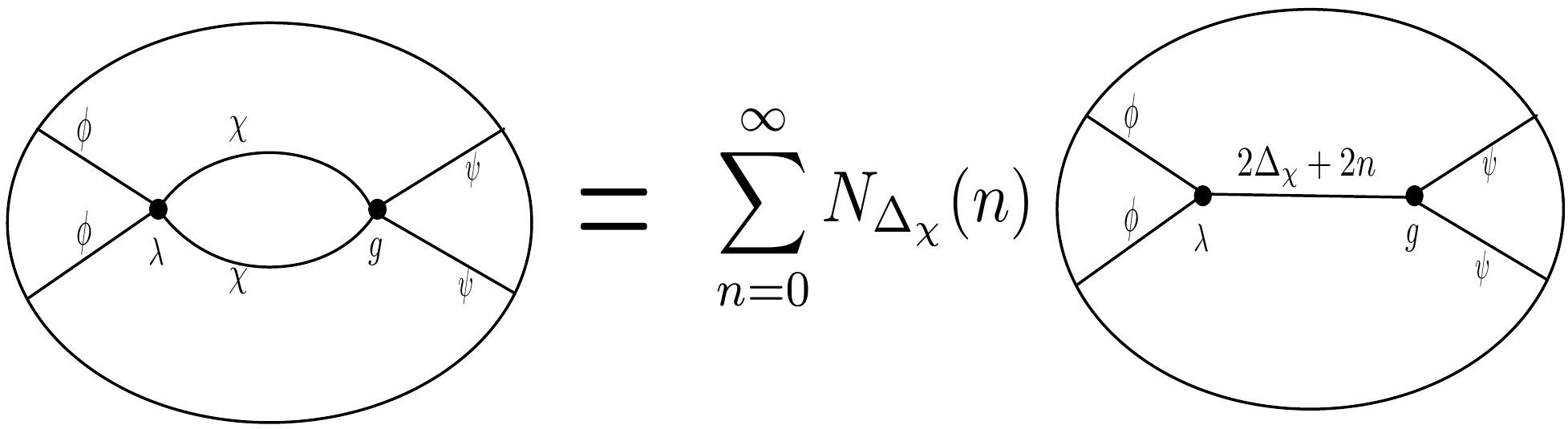}
\caption{ This figure shows how the one-loop example studied in section \ref{sec:OneLoopExample} can be computed by using the Kallen-Lehmann representation in the bulk of AdS, as discussed in \cite{Fitzpatrick:2011hu}.  In this way one can write the product of propagators in the loop as a sum over tree-level exchanges with different bulk masses, or CFT dimensions.  The conformal block decomposition of this Witten diagram was indicated in figure \ref{fig:WittenDiagramtoBlocks}.
\label{fig:KallenLehmannfor1Loop}  }
\end{center}
\end{figure}

Now let us put the pieces together and understand how the optical theorem applies to the one-loop amplitude for $2\phi \to 2 \psi$ in a theory with couplings $\frac{\lambda}{4} \phi^2 \chi^2$ and $\frac{g}{4} \chi^2 \psi^2$, as pictured in figure \ref{fig:KallenLehmannfor1Loop}.  We could just as easily treat $\phi^4$ theory, but we have introduced three fields in order to separate out the various different contributions to the one-loop amplitude.

We computed the relevant one-loop Mellin amplitude in \cite{Fitzpatrick:2011hu} and verified that it has the correct flat space limit.  We also explained how branch cuts arise from the coalescence of poles in the flat space limit, and verified that   the discontinuity across the branch cut is correctly reproduced.  Let us now summarize the method and results.  To compute a certain class of loop amplitudes, one can use the fact that in position space in the bulk of AdS 
\ba \label{eqn:ProductOfPropagators}
G_{\Delta_1}(X,Y) G_{\Delta_2}(X,Y) &=& \sum_n  a_{\Delta_1, \Delta_2}(n) G_{\Delta_1 + \Delta_2+2n}(X,Y),  \ \ \mathrm{where} \\
a_{\Delta_1,\Delta_2}(n) &=&  \frac{(h)_n}{2 \pi^h n! }  \frac{  (\Delta_1 + \Delta_2 + 2n)_{1-h}  (\Delta_1 + \Delta_2 + n -2h +1)_n }
{(\Delta_1 + n)_{1-h} (\Delta_2 + n)_{1-h} (\Delta_1 + \Delta_2 + n -h)_n }. \nonumber
\ea
This allows a Kallen-Lehmann type representation \cite{Dusedau:1985ue} for the loop amplitude as a sum over tree level exchanges with dimensions $2 \Delta_\chi + 2n$, as indicated in figure \ref{fig:KallenLehmannfor1Loop}.  Taking $N_{\Delta_\chi}(n) =   a_{\Delta_\chi,\Delta_\chi}(n)$ and using the diagrammatic rules from \cite{Fitzpatrick:2011hu} we find the Mellin amplitude for this diagram is
\be \label{eq:OneLoopExampleMellinAmp}
M^{\mathrm{1-loop}}(\delta_{ij}) = \lambda g \sum_n N_{\Delta_\chi}(n) M_{2 \Delta_\chi + 2n}(\delta_{ij})
\ee
where $M_\Delta(\delta_{ij})$ is the Mellin amplitude for a tree level exchange of a bulk field dual to an operator of dimension $\Delta$.  It is
\ba
M_{\Delta}(\delta_{ij}) &=& \sum_m \frac{R_m}{\delta - (\Delta + 2m)} ,
\ea
where $\delta =  2 \Delta_\phi  - 2 \delta_{12}$ and $\Delta = 2 \Delta_\chi + 2n$ in our case.  The residue $R_m$ is  \cite{Penedones:2010ue, Fitzpatrick:2011ia, Fitzpatrick:2011hu}
\be
R_m = -\frac{1}{(4\pi^h)^3 }\frac{\Gamma(\Delta_{\phi} - h + \frac{\Delta}{2})\Gamma(\Delta_{\psi}-h  + \frac{\Delta}{2})}{ \Gamma(\Delta_\phi - h + 1)^2 \Gamma(\Delta_\psi - h + 1)^2} \times \frac{\left(1 - \Delta_{\phi} +  \frac{\Delta}{2} \right)_m \left(1 - \Delta_{\psi} +  \frac{\Delta}{2}\right)_m}{ m!  \Gamma(\Delta - h + 1 + m) }
\label{eq:residues}
\ee
and it can be easily computed using the diagrammatic rules from \cite{Fitzpatrick:2011ia, Paulos:2011ie, Fitzpatrick:2011hu, Nandan:2011wc}.  

The particular decomposition in equation (\ref{eq:OneLoopExampleMellinAmp}) is very useful for several reasons.  First of all, to compute the flat space limit of the loop amplitude we need only know the flat space limit of $M_\Delta(\delta_{ij})$.  But this is simply the flat space scattering amplitude corresponding to the tree-level exchange of a particle of mass $(2 \Delta_\chi + 2n)/R$.  In this limit, the sum over $n$ becomes an integral and we can take the large $n$ limit of $N_{\Delta_\chi}(n)$ to find the loop amplitude
\be \label{eqn:1LoopFlatSpacePhi4}
\mathcal{M}^{\mathrm{1-loop}}(s) = \lambda g \int_0^\infty dn \frac{N_{\Delta_\chi}(n)}{s - (2 \Delta_\chi + 2n)^2} \ \ \ \ \mathrm{where}  \ \ \ \
N_{\Delta_\chi}(n) \approx
\frac{2}{(4 \pi )^{h} \Gamma (h) } 
n^{2(h-1)}
\ee
Another useful feature of equation (\ref{eq:OneLoopExampleMellinAmp}) is that it can be immediately related to the conformal block decomposition.  As we discussed in \cite{Fitzpatrick:2011hu}, conformal blocks and tree-level AdS exchanges have identical poles and residues in the Mellin amplitude; they differ only in their asymptotic behavior at large $\delta_{ij}$.  This means that we can immediately read off the coefficients in the conformal block decomposition of the correlator corresponding to this 1-loop Mellin amplitude in the $[\CO_\chi \CO_\chi]_{n,0}$ channel, it is
\ba
M_4(\delta_{ij}) &\supset & \sum_{n} P_{2 \Delta_\chi+ 2n} \times B_{2 \Delta_\chi + 2n,0}(\delta_{ij}) 
\ea
where
\ba \label{eq:OneLoopBlockCoeff}
P_{2 \Delta_\chi+ 2n} &=&  \lambda g N_{\Delta_\chi}(n) \left( \pi ^{\frac{1}{2}-3 h} 4^{-\Delta _{\chi }-n-2} \right) 
\\
&& \times \frac{ \Gamma
   \left(\frac{\Delta}{2} \right)^3 \Gamma \left(\Delta _{\psi } - \frac{\Delta}{2} \right) \Gamma \left(\Delta _{\phi } - \frac{\Delta}{2} \right) \Gamma \left(\Delta _{\psi } + \frac{\Delta}{2} -h \right)
   \Gamma \left(\Delta _{\phi } + \frac{\Delta}{2} -h \right)}{\Gamma
   \left(\Delta _{\psi }+1-h\right)^2 \Gamma \left(\Delta _{\phi
   }+1 - h\right)^2 \Gamma \left(\frac{\Delta + 1}{2}\right)
   \Gamma \left(\Delta+1 -h\right)} \nn
\ea
with $\Delta \equiv 2 \Delta_\chi + 2n$, and we have taken $\ell = 0$ because the coefficient of all the blocks with $\ell > 0$ vanish.  The normalization here follows from the relative definition of the Mellin amplitude for a bulk exchange and the normalization of the conformal blocks, which is most easily determined by relating the residues at their poles.  Finally, one can also see \cite{Fitzpatrick:2011hu} from equation (\ref{eqn:1LoopFlatSpacePhi4}) that the discontinuity across the branch cut in the flat space amplitude is just given by the residue of the pole in this equation, which is
\be
\mathrm{disc} \left[ M^{\mathrm{1-loop}} \right] = \lambda g \frac{N_{\Delta_\chi} \left(\frac{\sqrt{s}}{2} \right)}{4 \sqrt{s}}.  
\ee
This formula appears on the left-hand side of the optical theorem, as the imaginary part of the scattering amplitude.  It remains to see how this is reproduced by the right hand side of the optical theorem.

To compute the right-hand side of the optical theorem and check equality, we need to apply our conglomeration procedure to compute the OPE coefficients that determine $N_{\Delta_\chi}(n)$.  Fortunately, we did this calculation in section \ref{sec:ExampleConglomerationCalcs} and found the OPE coefficient $\delta c_{2 \Delta_\chi + 2n}^{\phi \phi}$ for the operator $[\CO_\chi \CO_\chi]_{n, 0}$ in the product $\CO_\phi \times \CO_\phi$.  One can verify using equation (\ref{eq:TreeLevelc3}) that 
\be \label{eq:OneLoopOPEBlockRelation}
\left( \delta c_{2 \Delta_\chi + 2n}^{\psi \psi} \right) \left( \delta c_{2 \Delta_\chi + 2n}^{\phi \phi} \right) = P_{2 \Delta_\chi+ 2n}
\ee
using the expression we computed in equation (\ref{eq:OneLoopBlockCoeff}).  This provides a very non-trivial test of the conglomeration technique.   We can also interpret the product of OPE coefficients in terms of a 2-particle phase space integral in the flat space limit. By definition, we have that
\be
c^{\phi \phi}_{n \ell J} T_{ \Delta_\phi, \Delta_\phi}^{n, \ell, J} = \left\langle \CO_\phi(P_1) \CO_\phi(P_2) \right| 2 \Delta_\chi + 2n , \ell, J \rangle
\ee
If we apply equation (\ref{eq:SingleParticleNorm}) to the $\CO_\phi$ operators, then we see that at large $n$, the constant $c^{\phi \phi}_{n \ell J} $ can be interpreted as a flat space scattering amplitude between two $\phi$ particles in plane wave states and a $2\chi$ particle state in the spherical wave state of equation (\ref{eq:TwoParticlePhaseSpace}).  So each OPE coefficient can be interpreted as a tree-level scattering amplitude for $2\phi \to 2 \chi$ and $2 \psi \to 2 \chi$, respectively, and in the flat space limit equation (\ref{eq:OneLoopOPEBlockRelation}) becomes the optical theorem.

\section{Discussion}
\label{sec:Discussion}

Although it has been clear for some time that the S-Matrix is the only exact observable in flat space quantum gravity, it has remained somewhat mysterious what sort of holographic theory \cite{Susskind:1994vu, ArkaniHamed:2008gz, ArkaniHamed:2009dn} might compute it.   A true theory should do more than just output scattering amplitudes, it must also provide an understanding of how fundamental principles such as bulk locality and quantum mechanics emerge.  This would appear to be an especially difficult problem if one tries to obtain a holographic theory that `lives' directly on the null boundaries of flat spacetime, where notions such as time and distance can only have a limited meaning.

We have argued that the flat spacetime limit of AdS/CFT may provide the theory that we have been looking for.  In \cite{Fitzpatrick:2011hu} we derived a formula conjectured by Penedones \cite{Penedones:2010ue} that relates the Mellin amplitude \cite{Mack, MackSummary} for $n$-pt CFT correlators to the flat space S-Matrix.  This formula expresses the S-Matrix as a simple integral transform of the Mellin amplitude, which itself must be a meromorphic function restricted to have only simple poles on the real axis.  The clearest way to understand flat spacetime locality from a holographic perspective is via the analyticity properties of the S-Matrix.  This strongly suggests that the most natural way to understand locality may be in terms of the very restricted analytic structure of the Mellin amplitude combined with some set of assumptions about the spectrum of the CFT, as described in \cite{JP}.

In the present work we have shown that the operator product expansion and conformal block decomposition of CFTs encodes unitarity in a form that appears very similar to the usual optical theorem for the S-Matrix, as pictured in figure \ref{fig:ConglomeratingtoBlocks}.    By taking the flat space limit of the AdS/CFT duality we showed that one can derive the optical theorem directly from these unitarity relations for the case of 2-to-2 scattering of massless scalar particles.  In the perturbative case this reduces to the usual cutting rules for Feynman diagrams.  We also saw something subtle and interesting occur with the spectrum -- in the flat space limit, CFT operators dual to unstable particles must decouple from the unitarity relation.  It would be interesting to understand this phenomenon better, and to explain it without having to appeal to bulk reasoning.   To give a vaguer and more ambitious-sounding summary,  one might say that we have shown how to derive  bulk quantum mechanics from the quantum mechanics of the holographic dual.

To make these derivations possible we developed technology to conglomerate many local operators together into a single  composite operator.  To enact this conglomeration we used smearing functions $f_{\Delta, \ell}$ that we labeled `wavefunctions' because, via the operator-state corresopndence, these functions also extract definite states from the CFT.  The relative simplicity of our formalism was made possible through use of the Mellin amplitude for CFT correlators.  Since the Mellin representation depends on position space kinematics through specific power-laws, it accords naturally with the structure of the conglomerating wavefunctions, which have the power-law behavior of CFT 3-pt correlators.  Perhaps in the future the logic will be reversed; one might attempt to derive the Mellin amplitude as the representation where these wavefunctions behave most naturally.

We also gave a schematic argument that the usual Dyson series for the S-Matrix can be constructed in AdS using the dilatation operator, and that at first order it gives rise to the prescription for the S-Matrix in terms of the anomalous dimension matrix that was given in \cite{Katz}.  In \cite{Fitzpatrick:2011hu} we derived Penedones formula for the S-Matrix using essentially the same wavepacket setup that was originally described in \cite{Polchinski, Susskind} and further examined in \cite{GGP, Fitzpatrick:2011jn}.  Thus we have given at least a rough explanation why all three of these holographic formalisms for computing the bulk S-Matrix agree, although we view the formula in terms of the Mellin amplitude as by far the most elegant and physical.

The impediment to formulating a general proof of the optical theorem for any $n$-pt scattering amplitude with particles of arbitrary spin seems to be mostly technical.  The requisite CFT technology to describe higher-spin particles \cite{Costa:2011mg, Costa:2011dw} and higher-point conformal blocks has not been fully developed, so we lack the necessary CFT ingredients.  Note that the fact that sums over $k$-trace operators reduce to $k$-particle phase space integrals is essentially kinematic, following from the structure of the conformal algebra when applied to large dimension operators.  Thus in the future one should be able to show how superconformal theories give rise to higher dimensional phase space integrals corresponding to extra dimensions that decompactify in the flat space limit. 

Perhaps it would be appropriate to mention here that some of the standard lore concerning $n$-pt correlators seems a bit misleading, and that this becomes very apparent when it is re-stated in terms of the bulk S-Matrix.  CFTs are often viewed as extremely constrained theories, and the statement is often made that all CFT correlators are completely determined by the 3-pt correlators.  In $2$-dimensions where there is an infinite dimensional symmetry algebra this may be a very powerful point, but in higher dimensions it is rather trivial.  In particular, the flat space limit of this statement is the trivial claim that once we know the $2$-to-$k$ scattering amplitudes for all $k$, we know the entire S-Matrix.  

We have mentioned this point in order to emphasize that higher correlation functions in CFTs are very non-trivial and are worthy of investigation.  Higher-point conformal blocks have rarely been discussed, but we would require such objects in order to formulate S-Matrix unitarity beyond four particles.  It would be interesting to develop techniques for handling them.  Another reason to study higher-point correlators is to understand Hawking radiation, which produces large multiplicity final states that can be understood in terms of the properties of correlators involving a large number of single-trace operators.  The flat space limit of AdS/CFT turns the bootstrap program for CFTs into the S-Matrix program, so focusing solely on the 4-pt correlators of single-trace operators would be like restricting the study of scattering amplitudes to 2-to-2 processes.
We hope that our portrait of the holographic S-Matrix suggests new directions for future investigation.

The AdS/CFT correspondence \cite{Maldacena, Witten, GKP} has proved to be an extremely important and fruitful discovery.  One perspective on this correspondence views it as a very large compendium of exact dualities between various CFTs and other models of quantum gravity, often in the form of Superstring Theories or M-Theory.   A complementary perspective involves  placing an arbitrary effective field theory in Anti-deSitter space and computing the correlation functions of bulk fields as they approach the boundary \cite{Katz, Harlow:2011ke}.  The correlators of this `effective conformal theory' will approximate those of a CFT to very good accuracy \cite{JP}, with errors suppressed by powers of operator dimensions divided by the cutoff in AdS units \cite{Katz}.  The search for a non-perturbative completion to a gravitational EFT in AdS space translates into the question of whether there exists an exactly defined CFT that approximately reproduces the boundary correlators of the AdS effective theory.  Similarly, we can UV complete a gravitational EFT in flat spacetime if we can find a sequence of CFTs of increasing central charge whose spectrum and correlators approximate those of the EFT in the flat space limit of AdS. 

In other words, although it may be extremely challenging to actually find a non-perturbative completion for a given gravitational effective field theory, anyone can use AdS/CFT to correctly formulate the question.  This means that one can obtain robust results about quantum gravity by modeling the correlators of low-dimension operators using bulk effective field theory, and then using the bootstrap approach \cite{Belavin:1984vu, Dolan:2003hv, Rattazzi:2008pe, Poland:2010wg, Costa:2011mg, Costa:2011dw, Hellerman:2009bu, Poland:2011ey, Rychkov:2011et,Fitzpatrick:2011hh} to constrain correlators that involve operators of larger dimension.  Since dimensions in the CFT correspond to bulk energies, one can obtain information about processes at trans-Planckian energies in AdS.  In \cite{Fitzpatrick:2011hu} we used Hawking evaporation to make a prediction for the conformal block decomposition of 4-pt correlators.  If one can derive this generic behavior from CFT dynamics by using the bootstrap, then quantum gravity may be accessible to mortals.

Should we view a holographic description as the final word on quantum gravity in a particular class of spacetimes?  The legalistic answer may be yes, but it seems that holographic descriptions such as AdS/CFT do not readily yield information about the physics behind horizons, and we might hope that such questions are not entirely ill-defined.  It seems reasonable to assume that if we could do experiments on black holes in a large enough laboratory, we would see unitary evaporation, with intrinsic errors that decrease as we increase the size of our detector.  This suggests that it may be worth looking \cite{Banks:2011av} for an approximation scheme beyond effective field theory that encodes both locality and its demise.

\section*{Acknowledgments}

We would like to thank Nima Arkani-Hamed, Tom Banks, Raphael Bousso, Daniel Harlow, Ami Katz, Dhritiman Nandan, Miguel Paulos, Jo\~ao Penedones, Steve Shenker, David Simmons-Duffin, and Alessandro Vichi for discussions.   ALF was partially supported by ERC grant BSMOXFORD no. 228169.  JK acknowledges support from the US DOE under contract no.~DE-AC02-76SF00515.

\appendix

\section{Conglomerating Operators: Regularization Details}
\label{sec:congscalar}

Here, we will describe in more detail our regularization procedure for conglomerating operators.  Let us return to eq. (\ref{eq:scalarc2}), but keeping $\Delta$ arbitrary:

\ba
\bar c^{12}_{\Delta, 0} N^f_{\Delta, 0}=
\pi^{2h} \frac{\Gamma (\frac{\Delta_1 +\Delta_2 - \Delta}{2}) \Gamma \left(h-\Delta_1\right) \Gamma
   \left(h-\Delta_2\right) \Gamma
   \left(-h+\frac{\Delta+\Delta_1+\Delta_2}{2} \right)}{\Gamma
   (\Delta_1) \Gamma (\Delta_2) \Gamma
   \left(h+\frac{\Delta-\Delta_1 -\Delta_2}{2} \right) \Gamma (2h-\frac{\Delta+\Delta_1+\Delta_2}{2})} 
   \label{eq:ConglomeratingC12}
\ea

% \ba
% {\cal A} &=&\frac{c_3 }{P_{03}^{\frac{\Delta_1 + \omega - \Delta_2}{2}} P_{04}^{\frac{\Delta_2 + \omega - \Delta_1}{2}} P_{34}^{\frac{\Delta_1 + \Delta_2 - \omega}{2}}}\nn\\
% c_3 &=& A \pi^{2h} \frac{\Gamma (-n) \Gamma \left(\frac{d}{2}-\Delta_1\right) \Gamma
%   \left(\frac{d}{2}-\Delta_2\right) \Gamma
%   \left(-\frac{d}{2}+n+\Delta_1+\Delta_2\right)}{\Gamma
%   (\Delta_1) \Gamma (\Delta_2) \Gamma
%   \left(\frac{d}{2}+n\right) \Gamma (d-n-\Delta_1-\Delta_2)}\nn\\
%   \ea
% 
 To fix the normalization, we want to conglomerate operators again to obtain a two-point function.  As we will see in a moment, it is necessary to regulate by both taking
 the dimension $\Delta'$ of the second operator in the two-point function to be arbitrary and also to define it as a function of two positions $P_5$ and $P_{5'}$ that we will take to
 be equal in physical quantities.  That is, we will calculate the two-point function
 \be
 \< \CO_{\Delta,0}(P_0) \CO_{\Delta',0}(P_5, P_{5'})\>
 \ee
 where 
 \be
 \CO_{\Delta',0}(P_5, P_{5'}) =\frac{1}{N^f_{\Delta',0}}  \int d^d P_1 d^d P_2 \frac{1}{P_{12}^{\frac{2d-\Delta_1 - \Delta_2 - \Delta'}{2}} P_{15}^{\frac{\Delta_2 + \Delta'-\Delta_1}{2}} P_{25'}^{\frac{\Delta_1 + \Delta'-\Delta_2}{2}} } \CO_1(P_1) \CO_2(P_2)
 \label{eq:secondreg}
 \ee
 Then, combining (\ref{eq:scalarc2})  and (\ref{eq:secondreg}), a straightforward application of Symanzik's star formula demonstrates that
 \ba
 \< \CO_{\Delta,0} (P_0) \CO_{\Delta',0}(P_5, P_{5'}) \> &=& \int d^d P_1 d^d P_2 \frac{P_{12}^{\frac{ \Delta' + \Delta_1 + \Delta_2 - 2d}{2}} }{P_{15}^{\frac{\Delta_1 + \Delta'-\Delta_2}{2}} P_{25'}^{\frac{\Delta_2 + \Delta'-\Delta_1}{2}} }\left( \frac{\CC_{\Delta_1} \CC_{\Delta_2} P_{12}^{\frac{\Delta - \Delta_1 - \Delta_2}{2}}}{P_{01}^{\frac{\Delta+ \Delta_1 - \Delta_2}{2}} P_{02}^{\frac{\Delta+ \Delta_2 - \Delta_1}{2} }}\right) \nn \\
 &=& %A' c_3\frac{\pi^{2h} \Gamma(\hat{\delta}_{04} ) \Gamma(\hat{\delta}_{05}) \Gamma(\hat{\delta}_{45}) }{\Gamma(a_{35}) \Gamma(a_{34} -n) \Gamma(\Delta_1+n)} \frac{\Gamma(\hat{\hat{\delta}}_{05})\Gamma(\hat{\hat{\delta}}_{05'}) \Gamma(\hat{\hat{\delta}}_{55'}) }    { \Gamma(\Delta_2 + n + \hat{\delta}_{04}) \Gamma(a_{45})\Gamma(\hat{\delta}_{45})} \nn\\ && 
      \frac{\CC_{\Delta_1}\CC_{\Delta_2}}{P_{05'}^{\frac{\Delta+\Delta_1 - \Delta_2}{2}} P_{0 5}^{\frac{\Delta+\Delta_2-\Delta_1}{2}} P_{5 5'}^{\frac{\Delta'-\Delta}{2}}} , \nn\\
&& \times  \frac{\bar{c}^{12}_{\Delta,0}}{N_{\Delta',0}^f} \frac{\pi^{2h} \Gamma(\frac{\Delta'-\Delta}{2})\Gamma(h-\frac{\Delta+\Delta_2-\Delta_1}{2})\Gamma(h-\frac{\Delta+\Delta_1-\Delta_2}{2}) \Gamma(\Delta -h)  }{\Gamma(h)\Gamma(\frac{\Delta+\Delta_1 - \Delta_2}{2})\Gamma(\frac{\Delta+\Delta_2 - \Delta_1}{2})\Gamma(2h-\Delta)}.\nn\\
%\hat{\hat{\delta}}_{05} + \hat{\hat{\delta}}_{05'} &=& \Delta_2 + n + \hat{\delta}_{04} , \nn\\
%\hat{\hat{\delta}}_{05} + \hat{\hat{\delta}}_{55'} &=& \hat{\delta}_{45} , \nn\\
%\hat{\hat{\delta}}_{05'} + \hat{\hat{\delta}}_{55'} &=& a_{45}  .
\ea
where we have taken the limit $\Delta' \rightarrow \Delta$ in places where it does not produce singularities.  
Now, let us calculate the inner product of the states corresponding to $\CO_{\Delta,0}(P_0)$ and $\CO_{\Delta',0}(P_5,P_{5'})$ in the usual way in radial quantization by taking $x_0\rightarrow 0$ and $x_5, x_5'\rightarrow \infty$ and rescaling by the appropriate powers.  As we take $\Delta' \rightarrow \Delta$, this inner product has the interpretation of a normalization, so this fixes $N_{\Delta,0}^f$ as in eq.~(\ref{eq:SpinZeroNorm}), but with ``0'' replaced by $\frac{\Delta'-\Delta}{2}$:
\be
N^f_{\Delta, 0} = c^{12}_{\Delta,0} \frac{\pi^{2h} \Gamma(\frac{\Delta'-\Delta}{2})\Gamma \left(h- \frac{\Delta+\Delta_1-\Delta_2}{2} \right) \Gamma(\Delta -h) \Gamma \left(h- \frac{\Delta+\Delta_2-\Delta_1}{2} \right) }{\Gamma(h) \Gamma \left(\frac{\Delta+\Delta_1-\Delta_2}{2} \right) \Gamma \left(\frac{\Delta+\Delta_2-\Delta_1}{2} \right) \Gamma(d-\Delta)} .
\label{eq:ConglomeratingNorm}
\ee
Finally, putting together eq. (\ref{eq:ConglomeratingC12}) and (\ref{eq:ConglomeratingNorm}) to obtain the physical quantity $c_{\Delta,0}^{12}$, we may take $\Delta' = \Delta_1 + \Delta_2 + 2n$ and $\Delta=\Delta_1 + \Delta_2 + 2n + \epsilon$ and safely take the limit $\epsilon \rightarrow 0$, for which we obtain the finite result of eq. (\ref{eq:scalarOPEinfN}). 

\section{Conglomerating Operators: Spinning Conformal Blocks}
\label{sec:congspin}

We will extend the calculation in section \ref{sec:ConglomeratingOperators} to general spin.  The general form of three- and two-point functions
of primaries is
\ba
\langle [\CO_1 \CO_2]_{2n+\ell} (P_1) [\CO_1 \CO_2]_{2n+\ell}(P_2) \rangle &=& C_2^{2n+\ell}(-2)^{\ell} \frac{((Z_1 \cdot Z_2) (P_1 \cdot P_2) - (Z_2 \cdot P_1) (Z_1 \cdot P_2) )^\ell}{P_{12}^{\Delta_1 + \Delta_2 + 2n + 2 \ell}}, \nn\\
\langle \CO_1(P_1) \CO_2(P_2) [\CO_3 \CO_4]_{2n+\ell}(P_3) \rangle &=& C_3^{2n+\ell} \frac{((Z_3 \cdot P_1) P_{23} - (Z_3 \cdot P_2) P_{13} )^\ell}{P_{12}^{-n}
P_{23}^{\Delta_2 + n + \ell} P_{31}^{\Delta_1 + n + \ell}}.
\label{eq:univ2and3pt}
\ea
The general spin smearing functions are again just three-point functions with shadow fields: 
\ba
[\CO_1 \CO_2]_{n,\ell} (P_0) &=& A \int d P_1 d P_2 \frac{ ((Z_0 \cdot P_1) P_{20} - (Z_0 \cdot P_2) P_{10})^{\ell}}{P_{12}^{d-\Delta_1 - \Delta_2 - n} P_{02}^{\Delta_1 + n +\ell} P_{01}^{\Delta_2 + n + \ell}} \CO_1(P_1)
\CO_2(P_2) .
\ea

We will be interested in conglomerating $\CO_1$ and $\CO_2$ in order to obtain spinning operators in their OPE.  While this procedure works for extracting a general operator, it will be interesting to apply this to special cases as well.  In particular, our first application will be to the theory at infinite $N$, where the only operators that arise in the OPE are the double-trace operators.  

\subsection{Conglomerating Operators: Disconnected Four-point Function}

Since the only operators to consider in this case are the double-trace operators, we will label them by their indices $n, \ell$, where $\Delta= \Delta_1+ \Delta_2 + 2n +\ell$; i.e. we will use the notation $[\CO_1 \CO_2]_{n, \ell}$ rather than $[\CO_1 \CO_2]_{\Delta, \ell}$.  It will also be convenient to define $\Delta_a = \frac{\Delta_1 + \Delta_2}{2}$.      To extract the $N=\infty$ three-point function of $[\CO_1 \CO_2]_{n, \ell}$ with $\CO_1, \CO_2$, we integrate the smearing function against the disconnected diagram:
\ba
\< [ \CO_1 \CO_2]_{n, \ell}(P_0) \CO_1(P_3) \CO_2(P_4) \> &=&\frac{1}{N_{n , \ell}^f} \int d P_1 dP_2  \frac{ ((Z_0 \cdot P_1) P_{20} - (Z_0 \cdot P_2) P_{10})^{\ell}}{P_{12}^{d-2\Delta_a- n} P_{02}^{\Delta_1 + n +\ell} P_{01}^{\Delta_2 + n + \ell}} \frac{\CC_{\Delta_1} \CC_{\Delta_2}}{P_{13}^{\Delta_1} P_{24}^{\Delta_2} } . \nn\\ 
\label{eq:disconnectedspinningsmear}
\ea
To do this integral directly is more complicated than the ones we have encountered, because the contraction vector $Z_0$ for the indices of the spinning field acts like an additional point.
To perform this integral more simply we can use the fact \cite{Costa:2011mg, Costa:2011dw} that the three-point function of scalars with a spin-$\ell$ field
has to be made out of powers of the tensor 
\be
C_{0 AB} = Z_0^A P_0^B - Z_0^B P_0^A,
\ee
and the correlator can be written $\frac{(P_3 \cdot C_0 \cdot P_4)^{\ell}}{P_{03}^{a_{03}} P_{04}^{a_{04}} P_{34}^{a_{34}}}$. 
In particular, the correlation functions have the gauge symmetry $F(P_i, Z_i + \alpha_i P_i)= F(P_i, Z_i)$. 
Thus, one way to make this constraint manifest and simplify the calculation is just to pick a gauge.\footnote{  In order to gauge-fix the $Z_i$'s, we should first
decide what these abstract objects look like.  They are clearly not regular points like $P_i$'s in 
the boundary theory, because they satisfy $Z_i \cdot P_i=0$, but $Z_i \ne P_i$ (otherwise
the $C_i^{AB}$'s would vanish).  
Let us imagine for a moment however what they would look like if they were, so $Z_i$
projects down to the point $z_i$.  Then, $Z_i \cdot P_i=0$ implies
\ba
(z_i - x_i)^2 &=& 0 .
\ea
Obviously, in Euclidean space this requires $z_i=x_i$, which we do not want.  
Without loss of generality, let's take $x_i=0$.  Then, we need
\ba
& & z_i^2=0.
\ea
Now we see  that $z_i$ should simply be a point on the boundary 
with complexified coordinates.
So, in general, we need
\ba
z_i = x_i + q_i, \ \ \ \ \ q_i^2 = 0.
\ea
This also makes it very explicit why taking $Z_i$ to be one of the $P_j$'s in order to simplify the integrals
is not allowed.
We can work out how the gauge transformation $Z\rightarrow Z+\alpha P$ acts on $q$:
\ba
Z= (1,z^2, z^\mu)=(1,x\cdot(x+2q), x^\mu+q^\mu) &\stackrel{Z\rightarrow Z+\alpha P}{\longrightarrow}& (1+\alpha, x\cdot ((1+\alpha)x + 2q) , (1+\alpha)x^\mu + q^\mu) \nn\\
&\cong& (1, x\cdot (x+2\frac{q}{1+\alpha}), x^\mu + \frac{q^{\prime \mu}}{1+\alpha}).
\ea
In the last line, we have used the fact that $P$'s project down to boundary points by rescaling $P^+ \rightarrow 1$. Thus, the effect of the gauge transformation on $Z$ is
written in terms of $q$ very simply:
\ba
Z \rightarrow Z+\alpha P &\Leftrightarrow& q \rightarrow q' = \frac{q}{1+\alpha}.
\ea
}
Of course, we have to decide what gauge to choose.  Seemingly natural choices like
$Z_i \cdot P_i$ and $Z_i \cdot Z_i$ are not helpful, since they are already satisfied in any gauge.  The next best thing seems to be 
$Z_i \cdot P_j$ for one of the other $j$'s.  

How much does this simplify the computation of the OPE coefficient?  Let us take the gauge where $Z_0 \cdot P_4$ vanishes.
Now, any time we get contributions from a positive power of $Z_0 \cdot P_4$, we can drop them.  So far, this does
not use any knowledge of the final form of the correlation function other than the fact that it is gauge-invariant; for any
$Z_0$, we can always re-obtain the full gauge-invariant result by calculating in this gauge and then restoring
gauge invariance by taking $Z_0 \rightarrow Z_0 + \alpha P_0$, $\alpha = -\frac{Z_0 \cdot P_4}{P_0 \cdot P_4}$, 
which is simply the gauge transformation that took $Z_0 \cdot P_4=0$, and which clearly
reintroduces the gauge-invariant $P_3 \cdot C_0 \cdot P_4$. 

Having fixed a gauge, we can continue with the computation. The smearing integral to perform is
\ba
\left( -\frac{1}{2} \right)^{\ell} \sum_k \left( {\ell \atop k  } \right) (-1)^{\ell -k} \int d P_1 d P_2 \frac{ P_{1z}^k P_{2z}^{\ell -k} }{P_{12}^{d-2\Delta_a -n} P_{01}^{\Delta_2 +n +k}
P_{02}^{\Delta_1 + n + \ell -k} P_{13}^{\Delta_1} P_{24}^{\Delta_2}},
\ea
where $P_{iz} = -2 P_i \cdot Z_0$. Focusing  on individual terms, we compute
\ba
B_{\ell k} &\equiv &\int d P_1 d P_2 \frac{ P_{1z}^k P_{2z}^{\ell -k} }{P_{12}^{d-2\Delta_a -n} P_{01}^{\Delta_2 +n +k}
P_{02}^{\Delta_1 + n + \ell -k} P_{13}^{\Delta_1} P_{24}^{\Delta_2}} \nn\\ 
&=& \int \frac{d P_1  P_{1z}^k}{P_{13}^{\Delta_1} P_{01}^{\Delta_2 + n + k}} \int [d \delta_{0z} d \delta_{1z} d \delta_{4z}] \frac{\Gamma(-\delta_{0z} )\Gamma(-\delta_{1z}) \Gamma(-\delta_{4z}) 
P_{0z}^{\delta_{0z}} P_{1z}^{\delta_{1z}} P_{4z}^{\delta_{4z}}}{\Gamma(-(\ell-k))} \nn\\
   && (\Delta_1 + n + \ell - k )_{\delta_{0z}} (d-2\Delta_a -n)_{\delta_{1z}} (\Delta_2)_{\delta_{4z}} \nn\\
   &&  I(\Delta_1 + n + \ell - k + \delta_{0z}, d-2\Delta_a -n + \delta_{1z}, \Delta_2 + \delta_{4z}) ,
\ea
where we have introduced the notation
\ba
I(a, b, c) &\equiv& \int d^d P_2 \frac{1}{P_{02}^{a} P_{12}^{b} P_{24}^{c}} = \chi(a,b,c) E_{14,0}^{a} E_{04,1}^{b} E_{01,4}^{c}, \ \ \ \ \ 
  E_{ij,k} \equiv \left( \frac{P_{ij}}{P_{ik} P_{jk}} \right)^{\frac{1}{2}} , \nn\\
  \chi(a,b,c) &\equiv& \frac{\pi^h \Gamma\left( \frac{a + b-c }{2} \right) \Gamma\left( \frac{b +c - a}{2} \right) \Gamma\left( \frac{c + a -b}{2} \right)}
  {\Gamma(a)\Gamma(b) \Gamma(c) }.
\ea
The $\delta_{iz}$'s satisfy the constraint $  \delta_{0z} + \delta_{1z} + \delta_{4z}= \ell -k,$ and the arguments of the $I$ function get shifted since we
have eliminated 3 $\tilde{\delta}_{ij}$ variables by using the constraints, $\tilde{\delta}_{04}+\tilde{\delta}_{14} = \Delta_2 + \delta_{2z}$, etc. 
Now, the only poles that survive our gauge choice are $\delta_{0z}=\delta_{4z}=0$, so $\delta_{1z} = \ell -k$, and we have
\ba
B_{\ell k} 
&=& \int \frac{d P_1  P_{1z}^\ell  \chi (\Delta_1 + n + \ell - k , d-2\Delta_a -n +\ell -k, \Delta_2 )(d-2\Delta_a -n)_{\ell-k} }{P_{13}^{\Delta_1} P_{01}^{h+n+\ell}P_{04}^{\Delta_1 + \Delta_2 +n-h}P_{14}^{h-\Delta_1 -n}} \nn\\
 &=& \frac{\chi(\Delta_1 + n + \ell - k , d-2\Delta_a -n +\ell -k, \Delta_2 )\chi(h+n+\ell,h-\Delta_1 -n, \Delta_1 + \ell)}{P_{04}^{\Delta_2+n}P_{03}^{\Delta_1+n+\ell}P_{34}^{-n}} P_{3z}^\ell \nn\\
  && \times (d-2\Delta_a -n)_{\ell-k}(\Delta_1)_\ell\nn\\
\ea
Somewhat remarkably,  performing the sum over $k$ obtains a relatively simple result for the three-point function:
\ba
\< \CO_1(P_3) \CO_2(P_4) [\CO_1 \CO_2]_{2n+\ell}(P_0) \> &=& \bar{c}^{12}_{n,\ell} \frac{ (P_3 \cdot Z_0)^\ell}{P_{04}^{\Delta_2+n}P_{03}^{\Delta_1+n+\ell}P_{34}^{-n}}, \nn\\
\bar{c}^{12}_{n,\ell} &=&
\frac{\CC_{\Delta_1} \CC_{\Delta_2}}{N_{n,\ell}^f} \frac{\pi^{2h} \Gamma(-n)\Gamma(h-\Delta_1) \Gamma(h-\Delta_2) \Gamma(-h+2\Delta_a + n + \ell)}
{\Gamma(h+n+\ell)\Gamma(\Delta_1) \Gamma(\Delta_2) \Gamma(2h-2\Delta_a -n)} .\nn\\
\label{eq:discthreepointspinning}
\ea
This indeed is the correct form of the three-point function in $Z_0 \cdot P_4=0$ gauge. 

Having obtained the three-point function, we next need to conglomerate again in order to determine the two-point function and thus the normalization of
$[\CO_1 \CO_2]_{2n+\ell}$.    The smearing integral we have to compute is
\ba
\< [\CO_1 \CO_2]_{2n+\ell}(P_0) [\CO_1 \CO_2]_{2n'+\ell}(P_5)\> &=& \frac{\bar{c}^{12}_{n,\ell}}{N_{n,\ell}^f} \int dP_3 dP_4 \frac{(Z_5 \cdot P_3 P_{45'} - Z_5 \cdot P_4 P_{35} )^\ell}{P_{34}^{d-2\Delta_a -n'} P_{35}^{\Delta_2 + n' + \ell}
P_{45'}^{\Delta_1 + n' +\ell} } \nn\\
  && \times \frac{(Z_0 \cdot P_3 P_{04} - Z_0 \cdot P_4 P_{03})^\ell}{P_{04}^{\Delta_2 + n + \ell} P_{03}^{\Delta_1 + n + \ell} P_{34}^{-n}},
\ea
where we have done the usual $n\ne n', 5\ne 5'$ regularizations. We will take the gauge $P_0 \cdot Z_5=0, P_5 \cdot Z_0 = P_{5'} \cdot Z_0 =0$ (the second two are the same because
$P_5=P_{5'}$ everywhere except when they may give rise to $P_{55'}$).  Binomially expanding, we have
\ba
\< [\CO_1 \CO_2]_{2n+\ell}(P_0) [\CO_1 \CO_2]_{2n'+\ell}(P_5)\> &=& \left( -\frac{1}{2} \right)^{2\ell} \frac{\bar{c}_{n,\ell}^{12}}{N_{n,\ell}^f} \sum_k^\ell \sum_{k'}^\ell (-1)^{k+k'} \left( { \ell \atop k} \right) \left( {\ell \atop k'} \right) B_{\ell kk'}, \nn\\
 B_{\ell k k'} &=& \int d P_3 d P_4 \frac{P_{3 z'}^{k'} P_{4 z'}^{\ell -k'} P_{3z}^k P_{4z}^{\ell -k}}{P_{34}^{d-2\Delta_a -n'-n } P_{35}^{\Delta_2 + n' + k'} P_{45'}^{\Delta_1 + n' + \ell - k'} P_{04}^{\Delta_2+n+\ell-k}
 P_{03}^{\Delta_1 + n + k}}. \nn\\
 \ea
 We first perform the $P_4$ integration, which results in the introduction of five new $\delta$ integration variables after imposing constraints.  Four of them are forced to vanish by our gauge choice, leaving a single variable that can be converted by the residue theorem into a sum:
 \ba
B_{\ell k k'} &=& \sum_m \frac{1}{P_{05'}^{-h+2\Delta_a +n+n'+m}}  \int dP_3 \frac{P_{3z'}^{\ell-m} P_{3z}^{\ell-m} P_{z z'}^m }{P_{35}^{\Delta_2+n'+k'} P_{03}^{\Delta_1+n+k}} 
(-1)^{m+1} \frac{(\ell-k)!(\ell-k')!}{m!(\ell-k-m)!(\ell-k'-m)!}\nn\\
 && \times \frac{ \chi(d-2\Delta_a - n' -n +2\ell-k-k'-2m, \Delta_1 + n' + \ell - k' , \Delta_2+n+\ell-k)  }
{P_{03}^{h-\Delta_1 -n'+\ell-k-m}P_{35'}^{h-\Delta_2 -n+\ell-k'-m}} \nn\\
 && \times  (d-2\Delta_a - n' -n)_{2\ell-k-k'-2m} .
 \ea
 We can simplify this by taking $P_{5'} \rightarrow P_5 $ in places where it will not lead to singularities, which in particular
 is any place that does not have a $n$ or $n'$ exponent.  So, we can exchange the
 powers of $P_{35}$ and $P_{35'}$ in the denominator for $P_{35}^{h+n'+\ell-m}P_{35'}^{-n}$.  But, then there are no
 powers of $k,k'$ remaining in the $P_{ij}$'s, so we can complete the sum over them outside the integral.  
 We can furthermore take $n\rightarrow n'$ in the prefactor, since this is needed as a regulator only in the powers of $P_{ij}$'s.  
 We thus obtain
 \ba
&& \< [\CO_1 \CO_2]_{2n+\ell}(P_0) [\CO_1 \CO_2]_{2n'+\ell}(P_5)\> = \nn\\
 && \frac{\bar{c}^{12}_{n,\ell}}{(-2)^{2\ell}N_{n,\ell}^f} \sum_{m=0}^\infty \frac{\pi^h (-1)^{m+1}(\ell !)^2 \Gamma(h-n - \Delta_1)\Gamma(h-n-\Delta_2) \Gamma^2( \ell + 2n + 2\Delta_a-h)}{m! \Gamma^2(1+\ell-m) \Gamma(\Delta_1 + \ell + n)
  \Gamma(\Delta_2 + \ell + n) \Gamma(2h -2n -2\Delta_a) \Gamma( m + 2n + 2\Delta_a-h)} \nn\\
 &&  \ \ \ \ \ \times \int d P_3 \frac{P_{3z}^{\ell-m} P_{3z'}^{\ell-m} P_{zz'}^m }{ P_{03}^{h+n-n'+\ell-m} P_{35'}^{-n} P_{35}^{h+n'+\ell-m} P_{05'}^{-h+2\Delta_a + n + n' +m}}.
\ea 
This last line is exactly of the form of Symanzik's star formula:
\ba
 && \int d P_3 \frac{P_{3z}^{\ell-m} P_{3z'}^{\ell-m} P_{zz'}^m }{ P_{03}^{h+n-n'+\ell-m} P_{35'}^{-n} P_{35}^{h+n'+\ell-m} P_{05'}^{-h+2\Delta_a + n + n' +m}} \nn\\
 && \ \ \ \ \ \ =
 \frac{(-1)^{\ell+m+1} P_{z z'}^{\ell} (\ell-m)! \chi(h+n'+\ell-m, h+n-n'+\ell-m, -n)}{P_{05}^{2\Delta_a + 2n' +\ell} P_{5 5'}^{n'-n}},
 \ea
 where we have taken $P_{5'} \rightarrow P_5$ except inside $P_{55'}$ as well as $n\rightarrow n'$ in some places that do not lead to singularities.
 Putting everything together, we thus have
  \ba
&& \< [\CO_1 \CO_2]_{2n+\ell}(P_0) [\CO_1 \CO_2]_{2n+\ell}(P_5)\> = c_2^{n,\ell} \left( -\frac{1}{2} \right)^{2\ell} \frac{P_{z z'}^\ell}{ P_{05}^{2\Delta_a + 2n+\ell}}, \nn\\
&& c_2^{n,\ell} =\frac{\bar{c}_{n,\ell}^{12}}{N_{n,\ell}^f}
   \frac{\pi ^{2 h}\Gamma(0) (-1)^l l! \Gamma (h-n-\Delta_1) \Gamma
   (h-n-\Delta_2) \Gamma (2 l+2 n+2\Delta_a-1) \Gamma (-h+l+2 n+2\Delta_a)}{\Gamma (h+l) \Gamma (l+n+\Delta_1) \Gamma (l+n+\Delta_2) \Gamma (2 h-2
   n-2\Delta_a) \Gamma (l+2
   n+2\Delta_a-1)} .\nn\\
   \label{eq:disctwopointspinning}
\ea
From equations (\ref{eq:discthreepointspinning}) and (\ref{eq:disctwopointspinning}), we can choose the normalization factor $N_{n,\ell}^f$ to set a canonically normalized two-point function coefficient, $c_2^{n,\ell}=1$.  Then, we can read off the OPE coefficient, which after some simplification is
\ba
(\bar{c}^{12}_{n,\ell})^2 &=&
   \frac{(-1)^\ell \CC_{\Delta_1} \CC_{\Delta_2} (\Delta_1-h+1)_n (\Delta_2-h+1)_n
   (\Delta_1)_{\ell+n} (\Delta_2)_{\ell+n}}{\ell! n! (\ell+h)_n (\Delta_1+\Delta_2+n-2h+1)_n (\Delta_1+\Delta_2+2 n+\ell-1)_l (\Delta_1+\Delta_2+n+\ell-h)_n} . \nn\\
   \ea

\subsection{Conglomerating Operators: Connected Four-point Function}
\label{app:General4ptConglomeration}

Next, let us apply the conglomerating methods to connected four-point functions. Here, it is more convenient to work with the Mellin representation of the four-point function, since all position space integrations can be performed using Symanzik's start formula.  We will use as our Mellin coordinates the variables
\be
x= \Delta_a - \delta_{12} = \frac{\delta}{2}, \ \ \ \ \ \  y = \delta_{14} = \frac{\gamma+ \delta}{2},
\ee
since our expressions will typically be more compact in terms of these variables than in terms of $\delta_{ij}$'s or $\delta, \gamma$.  The four-point function can then be written as
\ba
\CA_4 &=& \int dx dy  \overbrace{M(x,y) \Gamma(\Delta_a -x) \Gamma(\Delta_b -x) \Gamma^2(x-y) \Gamma^2(y)}^{\tilde{M}(x,y)} \frac{1 }{P_{12}^{\Delta_a -x} P_{34}^{\Delta_b -x}
(P_{13} P_{24})^{x-y} (P_{14} P_{23})^y} , \nn\\
\ea
where $\tilde{M}(x,y)$ is the reduced Mellin amplitude.  
We obtain the OPE coefficient by smearing to produce a three-point function:
\ba
\CA_3 &=& \int dP_1 dP_2 \frac{ ((Z_0 \cdot P_1) P_{20} - (Z_0 \cdot P_2) P_{10} )^\ell }
{P_{12}^{d-2\Delta_a -n} P_{02}^{\Delta_a + n  + \ell} P_{01}^{\Delta_a + n + \ell} } \CA_4
\ea
We will again work in a gauge where $Z_0 \cdot P_4 =0$ in order to simplify the calculation.  In order to use Symanzik's formula, we will binomially expand to obtain
\ba
\CA_3 &=& \left( -\frac{1}{2} \right)^\ell \sum_k \left( \ell \atop k \right) (-1)^{\ell-k} B_{\ell k} \nn\\
B_{\ell k} &=& \int dx dy dP_1 dP_2 \frac{ \tilde{M}(x,y) P_{1z}^k P_{2z}^{\ell -k} } {P_{01}^{\Delta_a + n + k} P_{02}^{\Delta_a+n+\ell-k} P_{12}^{d-\Delta_a -n -x} P_{34}^{\Delta_b -x} (P_{13} P_{24})^{x-y} (P_{14} P_{23})^y} 
\ea
Performing the $P_1$ integration first, using Symanzik's formula, we obtain
\ba
B_{\ell k } &=& \sum_{m=0}^k \left( k \atop m \right)  B_{\ell m k} \nn\\
B_{\ell m k} &=& \int  [d\delta ] dx dy d P_2 \tilde{M}(x,y) \nn\\
 && \times \frac{P_{2z}^{\ell -m} P_{3z}^m}{P_{02}^{\delta_{02} + \Delta_a + n + \ell -k} P_{03}^{\delta_{03} } P_{04}^{\delta_{04}} P_{23}^{\delta_{23}+y}
 P_{24}^{\delta_{24}+x-y} P_{34}^{\delta_{34} + \Delta_b-x}} \nn\\
 && \times \pi^h \frac{\Gamma(\delta_{02}) \Gamma(\delta_{03}) \Gamma(\delta_{04}) \Gamma(\delta_{23}) \Gamma(\delta_{24}) \Gamma(\delta_{34})}
 {\Gamma(\Delta_a + n + k)\Gamma(d-\Delta_a -n -x)\Gamma(x-y) \Gamma(y) }.
 \ea
The $P_2$ integration also follows from the application of Symanzik's formula. The essential structure of the result of conglomerating is therefore that it introduces a projection function $H(\Delta, \ell; x,y)$:
\ba
\CA_3 &=&\frac{P_{3z}^\ell }{P_{03}^{\Delta_a+n+\ell}P_{04}^{\Delta_a + n} P_{34}^{\Delta_b -\Delta_a -n}}\int dx dy M(x,y) \Gamma(\Delta_a -x) \Gamma(\Delta_b -x) H(\Delta_a + n, \ell; x,y), \nn\\
\ea
where an important point is that $H(\Delta, \ell;x ,y)$ does not depend on $\Delta_a$ or $\Delta_b$:
\ba
&& H(\Delta, \ell; x, y ) = \pi^{2h} \left( -\frac{1}{2} \right)^\ell  \sum_{k=0}^\ell \sum_{m=0}^k(-1)^{\ell -k}  \left( \ell \atop k \right)  \left( k \atop m \right) 
 \Gamma (y)   \Gamma (x-y)\nn\\
 &&  \times \int d\delta_{04} d \delta_{34} \frac{\Gamma (\delta_{04}) \Gamma (\delta_{34})
  \Gamma (\Delta-\delta_{04}) \Gamma
   (y-\delta_{04}-\delta_{34}) \Gamma (x-\Delta-\delta_{34}) \Gamma (h+k-m-x+\delta_{34}) }{\Gamma (k+\Delta) \Gamma
   (x-\delta_{04}-\delta_{34}) \Gamma (d-\Delta-x)
   \Gamma (h-\Delta+\delta_{04}) \Gamma
   (h+l-m+\Delta-x+\delta_{34})} \nn\\
   && \times \Gamma
   (h-\Delta-y+\delta_{04}) \Gamma (h+l-m-x+\delta_{04}+\delta_{34}) \Gamma (-h+m+\Delta+x-\delta_{04}-\delta_{34}) .
\ea
We do not have a nice closed-form expression for $H(\Delta, \ell; x, y)$ in general like we do for the special case of $\int dy \Gamma(y)\Gamma(x-y) H(\Delta, 0; x,y)$ that appeared in section \ref{sec:ConglomeratingOperators}.  It is possible that such an expression exists and could be obtained with more effort, and could be useful for extracting OPE coefficients for specific theories.   In addition, one may perform the $\delta_{04}$ and $\delta_{34}$ integrations above using the residue theorem in order to obtain $H(\Delta, \ell; x,y)$ as a sum, which could perhaps be useful in some situations for numeric computations.  

\section{Double Trace Operators}
\label{app:DoubleTraceOperators}

The purpose of this appendix is to obtain a completely general recursion relation that expresses double trace primary operators in terms of linear combinations of derivatives acting on $\CO_1 \CO_2$.  The full conformal algebra is
\ba\label{eq:confalgebra}
\left[ M_{\mu\nu} , P_\rho \right] = i (\eta_{\mu\rho} P_\nu - \eta_{\nu\rho} P_\mu)
, && \left[ M_{\mu\nu} , K_\rho \right] = i (\eta_{\mu\rho} K_\nu  - \eta_{\nu\rho}
K_\mu) , \nn\\
\left[ M_{\mu\nu} , D \right] = 0 , && \left[ P_\mu, K_\nu \right] =
- 2 (\eta_{\mu\nu} D + i M_{\mu\nu} ) , \nn\\
\left[ D, P_\mu \right] =  P_\mu , && \left[ D, K_\mu \right] = - K_\mu .
\ea
Primary operators are those annihilated by the special conformal generator $K_\mu$, so that $[K_\mu, \CO] =0$.  We will work with operators that are eigenstates of $D$ and the angular momentum generators, so our double trace operators can be written as $[\CO_1 \CO_2]_{n, \ell}$.  In what follows we will generalize the computations of \cite{Mikhailov:2002bp, Penedones:2010ue}.

\subsection{Action of $K_\mu$ on $T(k,l-k,u_1, u_2, m)$}

By contracting with traceless symmetric polarizations $V$, we can write the action of $K_\mu$ on the double-trace operators 
in a particular basis.  Let us follow the notation of \cite{Penedones:2010ue} and write a general operator of the desired form as
\ba
T(k,l-k, u_1, u_2, m) &=& V^{\alpha_1 \dots \alpha_l} P_{\alpha_1} \dots P_{\alpha_k} P_{\mu_1} \dots P_{\mu_m} (P^2)^{u_1} \CO_1 
P_{\alpha_{k+1}} \dots P_{\alpha_l} P_{\mu_1} \dots P_{\mu_m} (P^2)^{u_2} \CO_2 . \nn\\
\ea
Then, the action of $K_\mu$ on this operator is as follows:
\ba
&& K_\mu T(k,l-k,u_1,u_2,m) =
   V\cdot \left[  2 u_1 (d-2u_1 -2 \Delta_1) P_\mu P_{\alpha_1} \dots P_{\alpha_k} P_{\mu_1} \dots P_{\mu_m} (P^2)^{u_1-1} \CO_1 (\dots) \CO_2  \right.
   \nn\\
   && \ \ \ \left.  -2 (\Delta_1 + m + k +2u_1 -1) \left(\sum_{\alpha_s} \eta_{\mu \alpha_s} \overbrace{P_{\alpha_1} \dots \hat{P}_{\alpha_s} \dots P_{\alpha_k} }^{P_{k-1}^{(s)}}\overbrace{P^{\mu_1} \dots P^{\mu_m}}^{P_m} \right. \right. \nn\\
    && \ \ \ \ \ \ \ \ \ \ \ \left. \left. + \sum_{\mu_s} \overbrace{P_{\alpha_1} \dots P_{\alpha_k}}^{P_k} \overbrace{P^{\mu_1} \dots \hat{P}^{\mu_s} \dots P^{\mu_m}  }^{P_{m-1}^{(s)}}\right)(P^2)^{u_1} \CO_1 (\dots) \CO_2 \right. \nn\\
    && \ \ \ \ \left. + 2 P_\mu \left( \sum_{s>r} \eta_{\alpha_s \alpha_r} P_{k-2}^{(s,r)} P_m + \sum_{s,r} \eta_{\alpha_s \mu_r} P_{k-1}^{(s)} P_{m-1}^{(r)} + \sum_{s>r} \eta_{\mu_s \mu_r} P_k P_{m-2}^{(s,r)} \right) (P^2)^{u_1} \CO_1 (\dots) \CO_2 \right. \nn\\
    && \ \ \ \ \left. + (1\leftrightarrow 2, k \leftrightarrow l-k)
   \right].
   \ea
   Here, we have defined $P_k^{(i)}, P_k^{(i,j)}$ as indicated, and $(\dots) \CO_2$ indicates the $\CO_2$ half of the double-trace operator before the $K_\mu$ action.  Performing the contractions with $V$ and symmetrizing, we can write the results in terms of 
\ba
T_\mu (k,l-k, u_1, u_2, m) &=& V_\mu^{\alpha_2 \dots \alpha_l} P_{\alpha_2} \dots P_{\alpha_k} P_{\mu_1} \dots P_{\mu_m} (P^2)^{u_1} \CO_1 
P_{\alpha_{k+1}} \dots P_{\alpha_l} P_{\mu_1} \dots P_{\mu_m} (P^2)^{u_2} \CO_2 , \nn\\
\ea
where $V_\mu$ is also symmetric and traceless.  Grouping like terms, we find
    \ba 
K_\mu T(k,l-k, u_1, u_2, m) =
   2u_1 (d-2 u_1 -2 \Delta_1) &\times& T_\mu(k+1, l-k, u_1 -1, u_2, m) \nn\\
   2u_2 (d-2 u_2 -2 \Delta_2) &\times& T_\mu(k, l-k+1, u_1, u_2-1, m) \nn\\
  - 2k (\Delta_1 + m + k + 2u_1-1)  &\times& T_\mu(k-1,l-k,u_1, u_2, m) \nn\\
  - 2(l-k) (\Delta_2 + m +l- k + 2u_2-1)  &\times& T_\mu(k,l-k-1,u_1, u_2, m) \nn\\
  -2m (\Delta_1 + m  + 2u_1-1)&\times&T_\mu(k,l-k+1,u_1,u_2, m-1)\nn\\
  -2m (\Delta_2 + m + 2u_2-1)&\times&T_\mu(k+1,l-k,u_1,u_2, m-1)\nn\\
    +m(m-1)&\times& T(k+1,l-k,u_1, u_2+1, m-2) \nn\\
    +m(m-1) &\times& T(k,l-k+1,u_1+1, u_2, m-2) . \nn\\
   \ea

Since we are demanding that the operator $[\CO_1 \CO_2]_{n, \ell} = \sum a(k_1, k_2, u_1, u_2, m) T(k_1, k_2, u_1, u_2, m)$, be primary, we obtain an equation for the coefficients:
\ba
0 &=& 2(u_1 + 1)(d-2u_1-2-2\Delta_1) a(k_1 - 1, k_2, u_1+1, u_2, m) \nn\\
 &&+2(u_2+1)(d-2 u_2 -2 -2\Delta_2) a(k_1, k_2-1, u_1, u_2+1, m) \nn\\
  && -2(k_1 + 1)(\Delta_1 +m+k_1 + 2u_1) a(k_1+1,k_2, u_1, u_2, m) \nn\\
   && -2(k_2 + 1)(\Delta_2 + m + k_2 + 2u_2) a(k_1,k_2+1,u_1, u_2, m) \nn\\
    && -2 (m+1) (\Delta_1+m +2 u_1) a(k_1, k_2-1,u_1, u_2, m+1) \nn\\
     && -2 (m+1) (\Delta_2 + m + 2u_2) a(k_1-1, k_2, u_1, u_2, m+1) \nn\\
    && + (m+1)(m+2) a(k_1-1, k_2, u_1, u_2-1, m+2) \nn\\
     && + (m+1)(m+2) a(k_1, k_2-1, u_1-1, u_2, m+2)
    \ea
Note that there are two terms here where the total spin has been incremented to $\ell + 1$, while in the remaining terms it has been decremented to $\ell -1$.  These two types of terms must cancel amongst themselves.  The two incremented terms imply the equation
\ba
0 &=& -2(k_1 + 1)(\Delta_1 +m+k_1 + 2u_1) a(k_1+1,k_2, u_1, u_2, m) \nn\\
   && -2(k_2 + 1)(\Delta_2 + m + k_2 + 2u_2) a(k_1,k_2+1,u_1, u_2, m) .
\ea
This is very constraining, since it completely fixes the $k$-dependence of $a$.  Making an ansatz
\ba
a(k, l- k, u_1, u_2, n-u_1-u_2) &=& 
s_{n,l}(k) b(u_1, u_2) 
\ea
we can solve for $s_{n,l}(k)$ uniquely:
\ba
s_{n,l}(k) &=& \frac{(-1)^k}{k!(l-k)! \Gamma(\Delta_1 +n +u_1 -u_2 + k)\Gamma(\Delta_2 + n +u_2 - u_1 +l -k) } .
\ea
Note that this agrees with the results of \cite{Penedones:2010ue} for the $n= 0$ case he computed.  Substituting this back into our constraint on $a$ for the spin $\ell - 1$ terms, we obtain an equation for $b(u_1, u_2)$.  After some simplification, this can be written
\ba
0&=& \frac{(n-u_1-u_2)(1+n-u_1-u_2) b(u_1-1,u_2) }{k(l-k + n-1 - u_1 + u_2 + \Delta_2)} 
+ \frac{(n-u_1 - u_2)(1+n-u_1-u_2) b(u_1, u_2-1) }{(k-l)(k+n-1 +u_1-u_2+\Delta_1) } \nn\\
   && - \frac{2 (n - u_1 - u_2)(n-1+u_1 -u_2 + \Delta_1) b(u_1, u_2)}{k(k+n-1+u_1-u_2 + \Delta_1) }
    + \frac{2(n-u_1 - u_2)(n-1-u_1 +u_2 + \Delta_2) b(u_1, u_2)}{(l-k) (l-k+n-1-u_1 + u_2 + \Delta_2)} \nn\\
       && - \frac{2(u_2+1)(-d+2(1+u_2+\Delta_2))b(u_1, u_2+1)}{k(l-k+n-1-u_1+u_2 + \Delta_2)} 
         + \frac{2(u_1+1)(d-2(1+u_1+\Delta_1))b(u_1+1,u_2)}{(k-l)(k+n-1+u_1-u_2+\Delta_1)} \nn \\
         \ea
Because of the $k$-dependence of $s(k)$, this has many terms that depend explicitly on $k$. But, we have just proven that
$b(u_1, u_2)$ cannot have any $k$-dependence! Thus, we can multiply through by all the terms in the denominators and
collect coefficients by powers of $k$.  This gives us exactly three equations for $b(u_1,u_2)$, one for each of $1,k,k^2$.
It is straightforward to check that one linear combination of these three equations vanishes, so in fact we obtain only two equations.
We can take linear combinations of the remaining two equations to obtain two recursion relations, one that  increments $u_1$ and another that increments $u_2$:
\ba
b(u_1+1,u_2) &=& \frac{m\left( (1+m) b(u_1, u_2-1) -2(l+m-1+2u_1 + \Delta_1) b(u_1, u_2) \right) }{2(u_1+1)(-d + 2(1+u_1 + \Delta_1))} \nn\\
b(u_1,u_2+1) &=& \frac{m\left( (1+m) b(u_1-1, u_2) -2(l+m-1+2u_2 + \Delta_2) b(u_1, u_2) \right) }{2 (u_2+1)(-d+2(1+u_2 + \Delta_2))}
\ea
where we define $m \equiv n-u_1 -u_2$  for concision.  This allows us to obtain the full solution for $a(k,l-k, u_1, u_2, m)$
for any $n,l$, up to a single overall normalization factor by beginning with $b(0,0)$ and then recursively increasing the parameters $u_1$ and $u_2$.  

\subsection{Solving in the Boundary Case}

Unlike in the case of $s_{n, \ell}(k)$, we are unaware of a full closed form solution for $b(u_1, u_2)$. However, the equations for $b(u_1, u_2)$ simplify when $u_2 = 0$ or $u_1=0$, giving 
\be
b(u_1+1,0) = \frac{-(m) (l+m-1+2u_1 + \Delta_1) b(u_1, 0)  }{2(u_1+1)(-h + (1+u_1) + \Delta_1 )}
\ee
and similarly with $1 \to 2$.  This has the simple solution
\be
b(u_1, 0) =(-1)^{u_1} \frac{(\Delta_1 + n + \ell -1)_{u_1} n! }{2^{u_1} \Gamma(u_1) \Gamma(n + 1 - u_1) (\Delta_1 - h)_{u_1}} 
\ee
when normalized so that $b(0,0) = 1$.

\subsection{Large $u,n$ behavior of the coefficients}

The solutions to the above recursion relations are seem to be very complicated in general.  However, we will look for simplifications at large $n$.  Our first step will be to obtain an recursion relation for just the diagonal elements, $u_1=u_2 \equiv u$.  This may be done by moving along the diagonal and near-diagonal $u_1 = u_2-1$, solving only for these elements and no others.  Formally, we can take $b(u,u) = b_d (u), b(u-1,u)=b_o(u)$; then, eliminating $b_o(u)$ will give us our recursion relation for the diagonal elements $b_d(u)$.  This may be easily done by shifting the argument where necessary, and we find
\ba
0&=&+u^2 b_d(u) (-(2 (\Delta_1+u)-d)) (2
   (\Delta_2+u)-d)-\frac{1}{4}b_d(u-2) (n-2 u+1)_4
\nn\\
 &+& \left(d (1-2 u)+\frac{1}{2} (2
  \Delta_1+2 l+2 n+2 u-4) (2\Delta_2+2 l+2 n+2 u-4) -2 (2 u-1)
   (l+n-1)+2 u^2\right)    \nn\\
   && \times \frac{1}{2} b_d(u-1) (n-2 u+1)_2 .
   \ea

 Now, we want to take the large $u,n$ limit of this.  At leading order in this limit, the shifts in $u$ make no difference and we simply find
 \ba
 0 &\approx& n^3 (3n-8u) b_d(u),
 \ea
 which implies that $b_d(u)$ is peaked around $u=\frac{3n}{8}$.  To go to higher orders, we can expand the shifts in $u$ as derivatives.
 We further take $u = \frac{3n}{8} + \delta$, with $\delta$ of $\CO(1)$.  The subleading behavior of the recursion relation is 
 \ba
 0 &\approx & n^3 \left(4 (23 + 3d +2l -2 (\Delta_1 + \Delta_2) -32 \delta) b_d(\delta) -5 n b_d'(\delta) \right).
 \ea
 This is a first-order ordinary differential equation, and is easily solved.  It has a simple approximate solution in terms of a Gaussian:
 \ba
 b_d (\delta) &\propto&  \exp\left( - \frac{(\delta-\delta_0)^2}{2 \sigma^2} \right) ,\nn\\
 \sigma^2 &=& \frac{5n}{128} , \ \ \ \ \ \ \ \ \delta_0 = \frac{23+3d +2l -2(\Delta_1 + \Delta_2)}{32} .
 \ea 
 Thus we see that these coefficients are sharply peaked around $u_1 = u_2 = \frac{3n}{8}$ for large $n$.

\bibliographystyle{utphys}
\bibliography{Mellinbib}

 \end{document}